\documentclass[a4paper,UKenglish]{lipics-v2016}
\pdfoutput=1 
\usepackage{microtype}

\usepackage{latexsym}
\usepackage{mathabx}
\usepackage{float}
\usepackage[scaled]{helvet}
\usepackage{stmaryrd}
\usepackage{alltt}
\usepackage{xspace}
\usepackage{verbatim}
\usepackage{wrapfig}
\usepackage[usenames,dvipsnames]{xcolor}
\hypersetup{linkcolor=black,citecolor=black,urlcolor=RubineRed}

\usepackage{pgf,tikz}  
\usepackage{cite}

\newcommand{\mute}[1]{\ifdefined\draftflag{#1} \else{} \fi}

\usepackage{skull}

\newcounter{ToDos}
\newcounter{WarnCounts}
\newcommand{\decorateWC}{
  \stepcounter{WarnCounts}
  \marginpar{\textcolor{red}{$\skull\ \theWarnCounts$}}}

\newcommand{\decorateTD}{
  \stepcounter{ToDos}
  \marginpar{\textcolor{red}{$\textbf{TO DO}_{\#\ \theToDos}$}}}

\newcommand{\signedComment}[3]
           {\mute{\textcolor{#2}{(#1: {#3})}\decorateWC}}

\newcommand{\todo}[1]{\mute{\textcolor{red}{(TO DO:{#1})}\decorateTD}}

\newcommand{\gad}[1]{\signedComment{GAD}{red}{#1}}

\newcommand{\etc}{\emph{etc}}
\newcommand{\ie}{\emph{i.e.}\xspace}

\newcommand{\eg}{\emph{e.g.}\xspace}

\newcommand{\etal}{\emph{et~al.}\xspace}

\newcommand{\dom}[1]{\mathsf{dom}(#1)}

\newcommand{\specK}[1]{\ensuremath{\textcolor{blue}{#1}}}

\newcommand{\act}[1]{\textsf{\small{#1}}}
\newcommand{\aux}[1]{\textit{#1}}

\newcommand{\esc}[1]{\text{\texttt{\small{#1}}}}
\newcommand{\kw}[1]{\text{\textbf{#1}}}

\newcommand{\Asgn}{\leftarrow} 

\newcommand{\dotcup}{\ensuremath{\mathaccent\cdot\cup}}

\newcommand{\lcl}{{\mathsf{s}}}
\newcommand{\env}{{\mathsf{o}}}
\newcommand{\joint}{{\mathsf{j}}}

\newcommand{\selfsub}{\mathsf{s}}
\newcommand{\othersub}{\mathsf{o}}
\newcommand{\jointsub}{\mathsf{j}}

\newcommand{\hist}{\chi} 
\newcommand{\histS}{\hist_{\, \selfsub}}
\newcommand{\histO}{\hist_{\, \othersub}}
\newcommand{\histJ}{\hist_{\, \jointsub}}
\newcommand{\hempty}{\emptyset}

\def\ordlist{\sigma}
\newcommand{\E}{\tau}
\newcommand{\C}{\kappa}

\newcommand{\tleq}{\mathrel{\leq_\ordlist}}
\newcommand{\tle}{\mathrel{<_\ordlist}}

\newcommand{\stableorder}{\Omega}
\newcommand{\prefix}[1]{-\,{\tleq}\,#1}

\def\ordlistP{\sigma'}
\newcommand{\stableorderP}{\stableorder'}

\newcommand{\tleP}{\mathrel{<_\ordlistP}}

\newcommand{\EP}{\tau'}
\newcommand{\CP}{\kappa'}
\newcommand{\histP}{\chi'} 
\newcommand{\histSP}{\hist_{\, \selfsub}'}
\newcommand{\histOP}{\hist_{\, \othersub}'}
\newcommand{\histJP}{\hist_{\, \jointsub}'}

\newcommand{\wppP}{W_p'}

\def\lgVy{\ensuremath{\mathsf{lastGY}}}

\newcommand{\wInit}{\mathsf{W_{Off}}}
\newcommand{\wWrite}{\mathsf{New}}
\newcommand{\wDirty}{\mathsf{Fwd}}
\newcommand{\wClean}{\mathsf{Done}}

\newcommand{\sOn}{\mathsf{S_{On}}}
\newcommand{\sOff}{\mathsf{S_{Off}}}

\makeatletter
\newcommand{\oset}[3][0ex]{%
  \mathrel{\mathop{#3}\limits^{
    \vbox to#1{\kern-3\ex@
    \hbox{$\scriptstyle#2$}\vss}}}}
\makeatother

\makeatletter
\newcommand{\ojset}[3][0ex]{%
  \mathrel{\mathop{#3}\limits^{
    \vbox to#1{\kern-5\ex@
    \hbox{$\scriptstyle#2$}\vss}}}}
\makeatother

\newcommand{\hunion}{\mathbin{\dotcup}}

\newcommand{\eqdef}{\mathrel{\:\widehat{=}\:}}
\newcommand{\hpts}{\mapsto}
\newcommand{\ldot}{\mathord{.}\,}

\def\FF{\mathsf{False}}
\def\TT{\mathsf{True}}

\newcommand{\tsPre}[1]{\ensuremath{{\textcolor{blue}{#1}}}}
\newcommand{\tsPos}[1]{\ensuremath{\textcolor{blue}{#1}}}
\newcommand{\logvar}[1]{\ensuremath{\textcolor{blue}{[#1].}}}

\newcommand{\var}[1]{\mathit{#1}} 
\newcommand{\num}[1]{{\text{{\scriptsize{#1}}}}}

\def\GYR{{\mathbf{{g^{+}}{y^{?}}{r^{*}}}}}
\def\RZ{{\mathbf{{{(g | y)^{+}}}{r^{*}}}}}

\def\lat{\langle}
\def\rat{\rangle}
\def\tbnd{\Asgn}
\newcommand{\actwrite}[2]{{#1}\,{:=}\,{#2}}

\title{Concurrent Data Structures Linked in Time}

\author[1,2]{Germ\'{a}n Andr\'{e}s Delbianco} 
\author[3]{Ilya Sergey}
\author[1]{Aleksandar Nanevski}
\author[1]{Anindya Banerjee}

\affil[1]{IMDEA Software Institute,  Madrid, Spain\\
  {\texttt{\{german.delbianco,
      aleks.nanevski,anindya.banerjee\}@imdea.org}}}

\affil[2]{Universidad Polit\'{e}cnica de Madrid, Spain}

\affil[3]{University College London, UK\\
  {\texttt{i.sergey@ucl.ac.uk}}}

\authorrunning{G.\,A. Delbianco and I. Sergey and A. Nanevski and %
  A. Banerjee}

\Copyright{Germ\'{a}n Andr\'{e}s Delbianco and Ilya Sergey and %
  Aleksandar Nanevski and Anindya Banerjee}

\keywords{Concurrent separation logics, Linearization points, Snapshot
  Algorithms, FCSL}

\theoremstyle{definition}
\newtheorem{proposition}[theorem]{Proposition}
\newtheorem{invariant}[theorem]{Invariant}

\begin{document}
\maketitle

\begin{abstract}

Arguments about correctness of a concurrent data structure are
typically carried out by using the notion of \emph{linearizability}
and specifying the linearization points of the data structure's
procedures.
Such arguments are often cumbersome as the linearization points'
position in time can be \emph{dynamic} (depend on the interference,
run-time values and events from the past, or even future),
\emph{non-local} (appear in procedures other than the one considered),
and whose position in the execution trace may only be determined after
the considered procedure has already terminated.

In this paper we propose a new method, based on a separation-style
logic, for reasoning about concurrent objects with such linearization
points. We embrace the dynamic nature of linearization points, and
encode it as part of the data structure's \emph{auxiliary state}, so
that it can be dynamically modified in place by auxiliary code, as
needed when some appropriate run-time event occurs.
We name the idea \emph{linking-in-time}, because it reduces temporal
reasoning to spatial reasoning. For example, modifying a temporal
position of a linearization point can be modeled similarly to a
pointer update in separation logic.
Furthermore, the auxiliary state provides a convenient way to
concisely express the properties essential for reasoning about clients
of such concurrent objects.
%
%
%
%
We illustrate the method by verifying (mechanically in Coq) an
intricate optimal snapshot algorithm due to Jayanti, as well as some
clients.

\end{abstract}

\section{Introduction}
\label{sc:intro} 
   
Formal verification of concurrent objects commonly requires reasoning
about linearizability~\cite{HerlihyW+TOPLAS90}. This is a standard
correctness criterion whereby a concurrent execution of an object's
procedures is proved equivalent, via a simulation argument, to some
sequential execution. The clients of the object can be verified under
the sequentiality assumption, rather than by inlining the procedures
and considering their interleavings. Linearizability is often
established by describing the \emph{linearization points} (LP) of the
object, which are points in time where procedures take place,
\emph{logically}.  In other words, even if the procedure physically
executes across a time interval, exhibiting its linearization point
enables one to pretend, for reasoning purposes, that it occurred
instantaneously (\ie, atomically); hence, an interleaved execution of
a number of procedures can be reduced to a sequence of atomic events.

Reasoning about linearization points can be tricky. Many times, a
linearization point of a procedure is not \emph{local}, but may appear
in another procedure or thread. Equally bad, linearization points'
place in time may not be determined statically, but may vary based on
the past, and even future, \emph{run-time} information, thus
complicating the simulation arguments. A particularly troublesome case
is when run-time information influences the logical order of a
procedure that has already terminated.
This paper presents a novel approach to specification of concurrent
objects, in which the dynamic and non-local aspects inherent to
linearizability can be represented in a procedure-local and
thread-local manner. 


The starting point of our idea is to realize what are the
shortcomings of linearizability as a canonical specification method
for concurrent objects.
Consider, for instance, the following two-threaded program manipulating
a correct implementation of stack by invoking its \texttt{push}
and \texttt{pop} methods, which are atomic, \ie, linearizable:
\begin{center}
\begin{tabular}{l || l}
\texttt{push(3);} & \texttt{push(4)}
\\
\texttt{t1 := pop(); } & \texttt{t2 := pop();}
\end{tabular} 
\end{center}
Assuming that the execution started in an empty stack, we would like
to derive that it returns an empty stack and \texttt{(t1, t2)} is
either \texttt{(3, 4)} or \texttt{(4, 3)}.
Linearizability of the stack guarantees that the overall trace of
\texttt{push}/\texttt{pop} calls is coherent with respect to a
sequential stack execution. However, it does not capture
\emph{client}-specific partial knowledge about the \emph{ordering} of
particular \texttt{push}/\texttt{pop} invocations in sub-threads,
which is what allows one to prove the desired result as a
composition of separately-derived partial specifications of the left and the right thread.

This thread-local information, necessary for compositional reasoning
about clients, can be captured in a form of \emph{auxiliary
  state}~\cite{OwickiG+CACM76} (a generalization of \emph{history
  variables}~\cite{AbadiL+lics88}), widely used in Hoare-style
specifications of concurrent
objects~\cite{SergeyNB+ESOP15,LeyWildN+POPL13,JungSSSTBD+POPL15,JungKBD+ICFP16}.
A testament of expressivity of Hoare-style logics for concurrency with
rich auxiliary state are the recent results in verification of
fine-grained data structures with helping~\cite{SergeyNB+ESOP15},
concurrent graph manipulations~\cite{SergeyNB+PLDI15},
barriers~\cite{JungKBD+ICFP16,DoddsJPSB+TOPLAS16}, and even
\emph{non-linearizable} concurrent objects~\cite{SergeyNBD+OOPSLA16}.

Although designed to capture information about events that happened
concurrently \emph{in the past} (hence the original name \emph{history
  variables}), auxiliary state is known to be of little use for
reasoning about data structures with \emph{speculative} executions, in
which the ordering of past events may depend on other events happening
in the \emph{future}. Handling such data structures requires
specialized metatheory~\cite{LiangF+PLDI13} that does not provide
convenient abstractions such as auxiliary state for client-side
proofs. This is one reason why the most expressive client-oriented
concurrency logics to date avoid reasoning about speculative data
structures altogether~\cite{JungKBD+ICFP16}.

\subsubsection*{Our contributions}

The surprising result we present in this paper is that by allowing
certain \emph{internal} (\ie,~not observable by clients) manipulations
with the auxiliary state, we can use an existing program logic for
concurrency, like, \eg,
FCSL~\cite{NanevskiLSD+ESOP14,SergeyNB+PLDI15}, to specify and verify
algorithms whose linearizability argument requires speculations, \ie,
depends on the \emph{dynamic reordering} of events based on run-time
information from the future.
%
%
To showcase this idea, we provide a new specification (spec) and the
first formal proof of a very sophisticated snapshot algorithm due to
Jayanti~\cite{Jayanti+STOC05}, whose linearizability proof exhibits
precisely such kind of dependence.

While we specify Jayanti's algorithm by means of a separation-style
logic, the spec nevertheless achieves the same general goals as
linearizability, combined with the benefits of compositional
Hoare-style reasoning.
In particular, our Hoare triple specs expose the logical atomicity of
Jayanti's methods (Section~\ref{sc:formal}), while hiding their true
fine-grained and physically non-atomic nature.  The approach also
enables that the separation logic reasoning is naturally applied to
clients (Section~\ref{sc:clients}).
%
%
Similarly to linearizability, our clients can reason out of
procedures' spec, not code. We can also ascribe the same spec to
different snapshot algorithms, without modifying client's code or
proof.


In more detail, our approach works as follows. 
We use shared auxiliary state to record, as a list of timed events
(\eg, writes occurring at a given time), the logical order in which
the object's procedures are perceived to execute, each instantaneously
(Section~\ref{sc:auxiliaries}). Tracking this time-related information
through state enables us to specify its dynamic aspects. We can use
\emph{auxiliary code} to mutate the logical order \emph{in place},
thereby permuting the logical sequencing of the procedures, as may be
needed when some run-time event occurs
(Sections~\ref{sc:implementation} and~\ref{sc:proof}). This mutation
is similar to updating pointers to reorder a linked list, except that
it is executed over auxiliary state storing time-related data, rather
than over real state. This is why we refer to the idea as
\emph{linking-in-time}.



Encoding temporal information by way of representing it as mutable
state allows us to use FCSL off-the-shelf to verify example
programs. In particular, FCSL has been implemented in the proof
assistant Coq, and we have fully mechanized the proof of Jayanti's
algorithm~\cite{CoqFiles}.

\newcommand{\fx}{\mathit{fx}}
\newcommand{\fy}{\mathit{fy}}
\newcommand{\x}{x}
\newcommand{\y}{y}
\newcommand{\s}{S}


\newcommand{\fwdp}[1]{(\esc{fwd}~#1)}
\newcommand{\aleksfwdp}[1]{\esc{fwd}~#1}

\begin{figure}[t]
\centering
\begin{tabular}[t]{l@{\ \ \ }l}
\begin{minipage}[t]{.4\textwidth}
\[
\begin{array}{rl}
\num{1}~ & \esc{write}\ (p,\, v)\ \{ \\ 
\num{2}~ & ~~~ \actwrite{p}{v}; \\
\num{3}~ & ~~~ b \tbnd \act{read}(\s);\\
\num{4}~ & ~~~ \kw{if}\ b\\
\num{5}~ & ~~~ \kw{then}\ \actwrite{\fwdp{p}}{v}\}\\
& \\
& \\
& \esc{fwd}\ (p : \esc{ptr})\ \{ \\
& ~~~ \kw{return}\ (p = x)\ \esc{?}\ \fx \esc{:}\ \fy\ \}
\end{array}
\]
\end{minipage}
&
\begin{minipage}[t]{.6\textwidth}
\[
\begin{array}{rl}
\num{6}~  & \esc{scan}\ :\ (A \times A)\ \{\\ 
\num{7}~  & ~~~~ \actwrite{\s}{\esc{true}};\\
\num{8}~  & ~~~~ \actwrite{\fx}{\bot};\\
\num{9}~  & ~~~~ \actwrite{\fy}{\bot}; \\
\num{10}~ & ~~~~ \var{vx} \tbnd \act{read}(\x);\\
\num{11}~ & ~~~~ \var{vy} \tbnd \act{read}(\y);\\
\num{12}~ & ~~~~ \actwrite{\s}{\esc{false}};\\
\num{13}~ & ~~~~ \var{ox} \tbnd \act{read}(\fx);\\
\num{14}~ & ~~~~ \var{oy} \tbnd \act{read}(\fy);\\
\num{15}~ & ~~~~ \var{rx} \tbnd \kw{if}\ (\var{ox} \neq\bot)\ \kw{then}\ \var{ox}\
                          \kw{else}\ \var{vx};\\  
\num{16}~ & ~~~~ \var{ry} \tbnd \kw{if}\ (\var{oy} \neq\bot)\ \kw{then}\ \var{oy}\
                          \kw{else}\ \var{vy};\\  
\num{17}~ & ~~~~ \kw{return}~(\var{rx},\var{ry})\}\\
\end{array}
\]
\end{minipage}
\end{tabular}
\caption{Jayanti's single-scanner/single-writer snapshot algorithm.}
\label{fig:jayanti-snapshot}
\end{figure}


\def\lineWrtStarts{1}
\def\lineWrtWrt{2}
\def\lineWrtChk{3}
\def\lineWrtIf{4}
\def\lineWrtFwd{5}
\def\lineWrtFnz{5'}

\def\lineScanStarts{6}
\def\lineScanSetsS{7}
\def\lineScanClearsX{8}
\def\lineScanClearsY{9}
\def\lineScanReadsX{10}
\def\lineScanReadsY{11}
\def\lineScanUnsetsS{12}
\def\lineScanReadsFX{13}
\def\lineScanReadsFY{14}
\def\lineScanChoosesRX{15}
\def\lineScanChoosesRY{16}
\def\lineScanRelinks{17}

\newcommand{\jywrite}{\texttt{write}\xspace}
\newcommand{\jyscan}{\texttt{scan}\xspace}

\section{Verification challenge and main ideas}
\label{sc:overview}

Jayanti's snapshot algorithm~\cite{Jayanti+STOC05} provides the
functionality of a shared array of size $m$, operated on by two
procedures: \jywrite, which stores a given value into an element, and
\jyscan, which returns the array's contents. We use the
\emph{single-writer}/\emph{single-scanner} version of the algorithm.
which assumes that at most one thread writes into an element, and at
most one thread invokes the scanner, at any given time. In other
words, there is a scanner lock and $m$ per-element locks. A thread
that wants to scan, has to acquire the scanner lock first, and a
thread that wants to write into element $i$ has to acquire the $i$-th
element lock. However, scanning and writing into different elements
can proceed concurrently.
%
This is the simplest of Jayanti's algorithms, but it already exhibits
linearization points of dynamic nature. We also restrict the array
size to $m\,{=}\,2$ (\ie, we consider two pointers $\x$ and $\y$,
instead of an array). This removes some tedium from verification, but
exhibits the same conceptual challenges.
 
The difficulty in this snapshot algorithm is ensuring that the scanner
returns the most recent snapshot of the memory. A na\"{i}ve scanner, which
simply reads $\x$ and $\y$ in succession, is unsound. To see why,
consider the following scenario, starting with $\x=5$, $\y=0$. The
scanner reads $\x$, but before it reads $\y$, another thread preempts
it, and changes $\x$ to $2$ and, subsequently, $\y$ to $1$. The
scanner continues to read $\y$, and returns $\x=5, \y=1$, which was
never the contents of the memory. Moreover, $(\x, \y)$, changed from
$(5,0)$ to $(2, 0)$ to $(2, 1)$ as a result of distinct
non-overlapping writes; thus, it is impossible to find a linearization
point for the scan because linearizability only permits reordering of
overlapping operations.



To ensure a sound snapshot, Jayanti's algorithm internally keeps
additional \emph{forwarding pointers} $\fx$ and $\fy$, and a boolean
\emph{scanner bit} $\s$. The implementation is given in
Figure~\ref{fig:jayanti-snapshot}.\footnote{Following Jayanti, we
  simplify the presentation and omit the locking code that ensures the
  single-writer/single-scanner setup. Of course, in our Coq
  development~\cite{CoqFiles}, we make the locking explicit.}
The intuition is as follows. A writer storing $v$ into $p$
(line~\lineWrtWrt), will additionally store $v$ into the forwarding
pointer for $p$ (line~\lineWrtFwd), provided $S$ is set. If the
scanner missed the write and instead read the old value of $p$
(lines~\lineScanReadsX--\lineScanReadsY), it will have a chance to
catch $v$ via the forwarding pointer
(lines~\lineScanReadsFX--\lineScanReadsFY). The scanner bit $S$ is
used by writers (line~\lineWrtChk) to detect a scan in progress, and
forward $v$.

{
\begin{figure}[t]
\captionsetup[subfigure]{justification=centering}
\centering  
\begin{subfigure}[t]{1\textwidth}
\centering
\begin{tabular}{l || l || l}
  \texttt{l: }\texttt{write (x,2);}\quad &
   \multirow{2}{*}{\texttt{c: scan ()}}\quad & 
    \multirow{2}{*}{\texttt{r: write (x,3)}}  \\
  \phantom{\texttt{l: }}\texttt{write (y,1)} & &   
\end{tabular}
\caption{\label{fig:weird:code}Parallel composition of three threads \texttt{l, c, r}.}
\end{subfigure}\\

\begin{subfigure}[b]{1\textwidth}
\begin{tabular}{l@{\hfill} l@{\hfil}}
\begin{minipage}[t]{0.5\textwidth}
\begin{alltt}
 \num{1}  c: \actwrite{S}{true}
 \num{2}  c: \actwrite{fx}{\(\bot\)}
 \num{3}  c: \actwrite{fy}{\(\bot\)}
 \num{4}  c: \act{read}(x)  // vx <- 5
 \num{5}  c: \act{read}(y)  // vy <- 0
 \num{6}  l: \actwrite{x}{2}
 \num{7}  l: \act{read}(S)  // b <- true
 \num{8}  l: \actwrite{fx}{2} 
 \num{9}  l: return ()
\num{10}  r: \actwrite{x}{3}
\end{alltt}
\end{minipage}
&
\begin{minipage}[t]{0.33\textwidth}
\begin{alltt}
\num{11} l: \actwrite{y}{1}
\num{12} l: \act{read}(S)  // b <- true
\num{13} l: \actwrite{fy}{1}
\num{14} l: return ()
\num{15} c: \actwrite{S}{false}
\num{16} r: \act{read}(S)  // b <- false
\num{17} r: return ()
\num{18} c: \act{read}(fx) // ox <- 2
\num{19} c: \act{read}(fy) // oy <- 1
\num{20} c: return (2,1)
\end{alltt} 
\end{minipage}
\end{tabular}
\caption{\label{fig:weird:exec} A possible interleaving of the threads
  in~(\subref{fig:weird:code}).}
\end{subfigure}
\caption{\label{fig:weird} An example leading to a scanner miss.%
}
\end{figure}
}

As Jayanti proves, this implementation \emph{is} linearizable. Informally,
every overlapping calls to \jywrite~and \jyscan~can be rearranged to
appear as if they occurred sequentially.  To illustrate, consider the
program in Figure~\ref{fig:weird:code}, and one possible interleaving
of its primitive memory operations in Figure~\ref{fig:weird:exec}. The
threads {\tt l}, {\tt c}, and {\tt r}, start with $\x = 5, \y = 0$.
The thread {\tt c} is scheduled first, and through lines~1--5 sets the
scanner bit, clears the forwarding pointers, and reads $\x = 5, \y =
0$. Then {\tt l} intervenes, and in lines~6--9, overwrites
$\x$ with $2$, and seeing $\s$ set, forwards $2$ to $\fx$. Next, {\tt
  r} and {\tt l} overlap, writing $3$ into $\x$ and $1$ into
$\y$. However, while $1$ gets forwarded to $\fy$ (line 13), $3$ is not
forwarded to $\fx$, because $\s$ was turned off in line 15 (\ie, the
scan is no longer in progress). Hence, when {\tt c} reads the
forwarded values (lines 18, 19), it returns $\x = 2, \y = 1$.

While $\x\,{=}\,2, \y\,{=}\,1$ was never the contents of the memory,
returning this snapshot is nevertheless justified because we can
\emph{pretend} that the scanner \emph{missed} {\tt r}'s write of
$3$. Specifically, the events in Figure~\ref{fig:weird:exec} can be
\emph{reordered} to represent the following sequential execution:
\begin{equation}
\hfill \mathtt{write\, (x, 2);\ write\, (y,1);\ scan\, ();\ write\, (x,
  3)}\hfill \label{eq:lin}
\end{equation}
Importantly, the client programs have no means to discover that a
different scheduling actually took place in real time, because they
can access the internal state of the algorithm only via interface
methods, \jywrite~and \jyscan.

This kind of temporal reordering is the most characteristic aspect of
linearizability proofs, which typically describe the reordering by
listing the linearization points of each procedure. At a linearization
point, the procedure's operations can be spliced into the execution
history as an uninterrupted chunk. For example, in Jayanti's proof,
the linearization point of \jyscan~is at line~\lineScanUnsetsS\ in
Figure~\ref{fig:jayanti-snapshot}, where the scanner bit is unset. The
linearization point of \jywrite, however, may vary. If
\jywrite~starts before an overlapping \jyscan's line~\lineScanUnsetsS,
and moreover, the \jyscan~misses the \jywrite---note the dynamic and
future-dependent nature of this property---, then \jywrite~should
appear after {\tt scan}; that is, the \jywrite's linearization point
is right after \jyscan's linearization point at line~\lineScanUnsetsS.
Otherwise, \jywrite's linearization point is at line~\lineWrtWrt.
In the former case, \jywrite~exactly has a non-local and
future-dependent linearization point, because the decision on the
logical order of this \jywrite~depends on the execution of \jyscan~in
a different thread. This decision takes effect on
lines~\lineScanReadsFX--\lineScanReadsFY, which can take place
\emph{after} the execution of \jywrite~has terminated.
For instance, in Figure~\ref{fig:weird:exec} the execution
of \jywrite~in \texttt{r} terminates at step 17, yet, in Jayanti's
proof, the decision to linearize this \jywrite\ after the
overlapping \jyscan\ is taken at line~18, when the \jyscan\ reads the
value from the previous \jywrite.




\begin{figure}[t]
\begin{subfigure}[t]{0.49\textwidth}
\includegraphics[width=6.1cm]{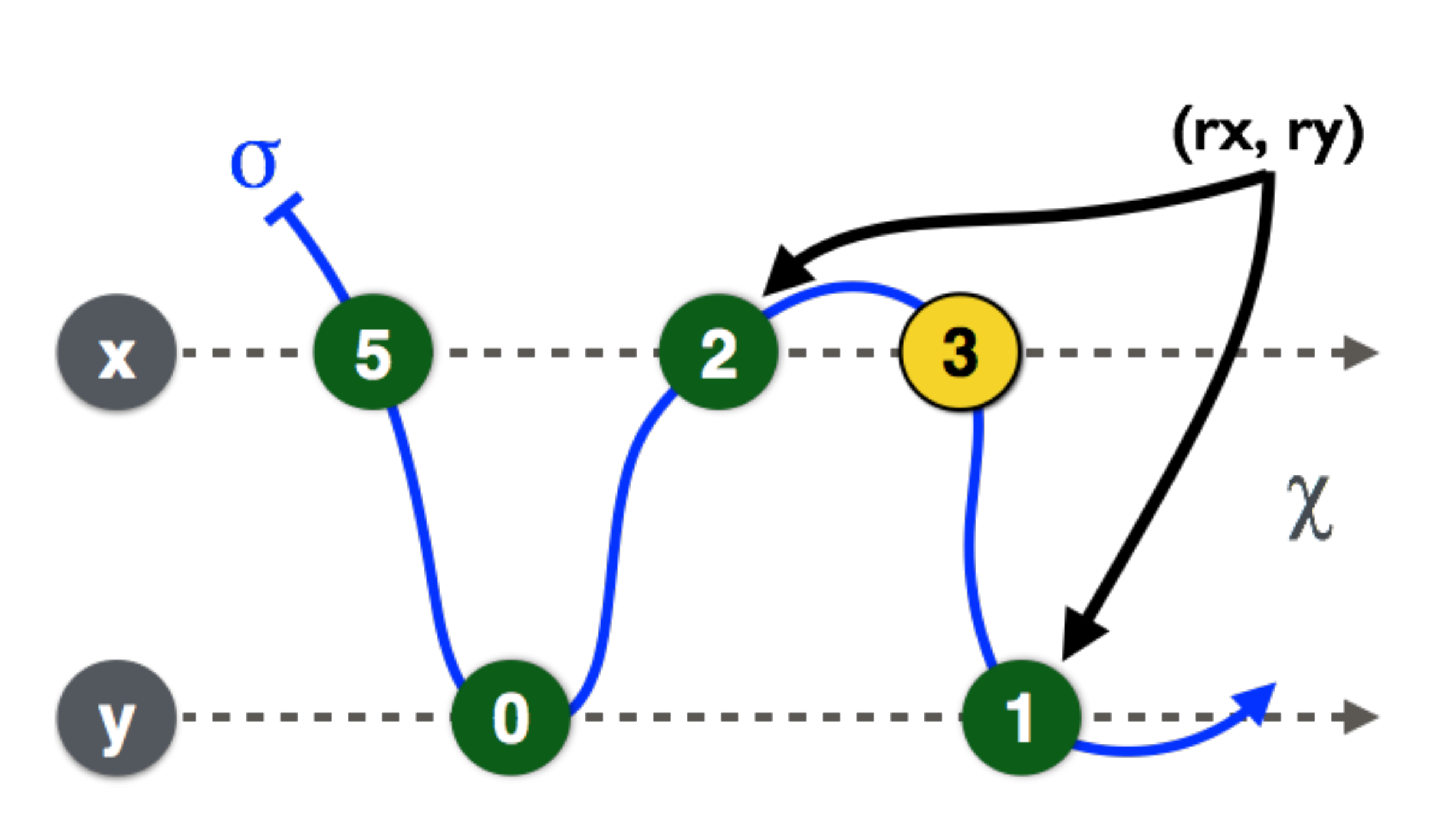}
\caption{\label{fig:reorder:before}} 
\end{subfigure} \hfill
\begin{subfigure}[t]{0.49\textwidth}
\includegraphics[width=6.1cm]{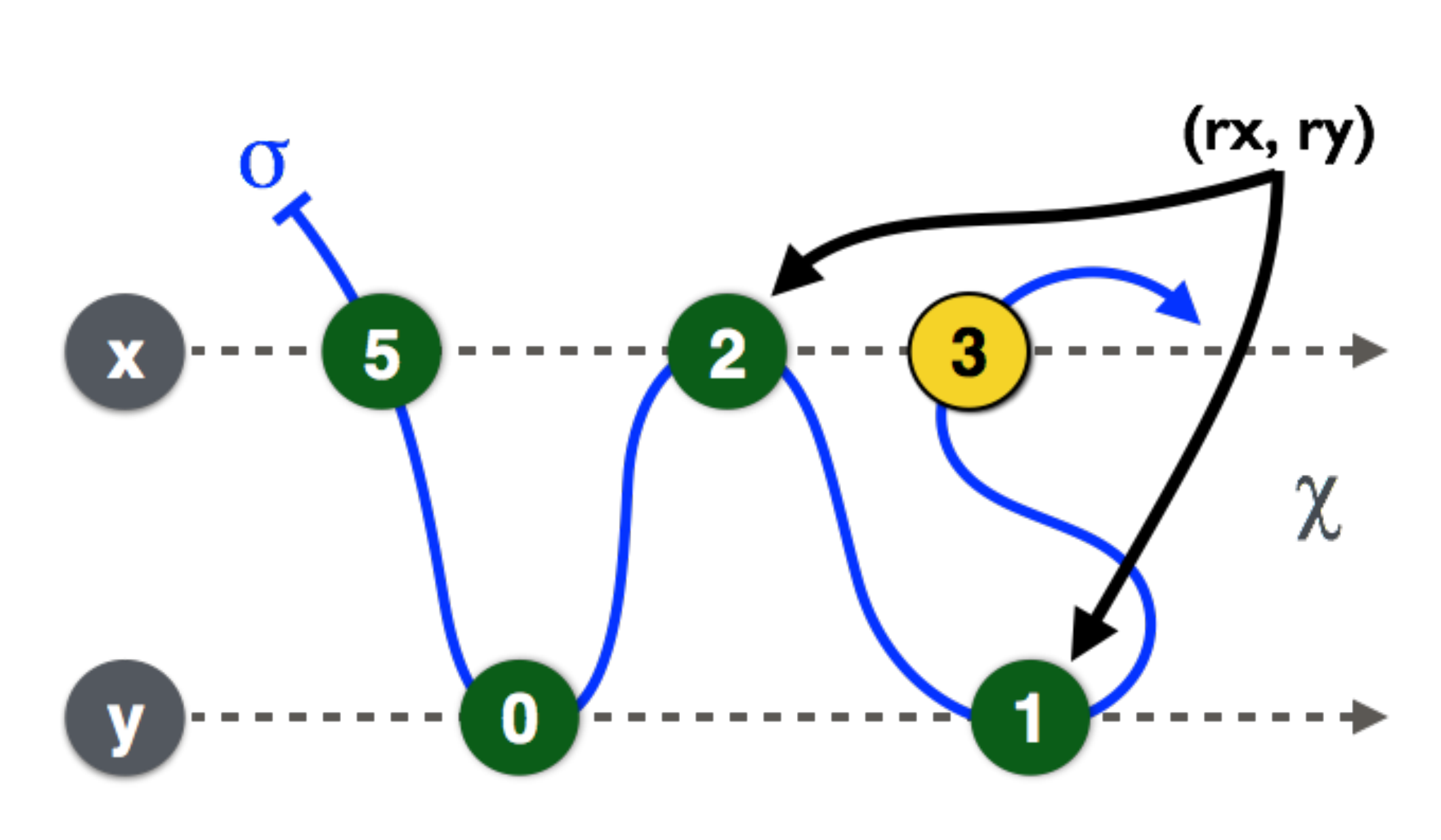}
\caption{\label{fig:reorder:after}} 
\end{subfigure}%
\caption{\label{fig:reorder} Changing the logical ordering (solid line
  $\ordlist$) of write events from (5, 0, 2, 3, 1) in
  (\subref{fig:reorder:before}) to (5, 0, 2, 1, 3) in
  (\subref{fig:reorder:after}), to reconcile with {\tt scan} returning
  the snapshot $\x=2, \y=1$, upon missing the write of $3$. Dashed
  lines $\hist$ represent real-time ordering.}
\end{figure}

Obviously, the high-level pattern of the proof requires tracking the
\emph{logical ordering} of the \jywrite\ and \jyscan\ events, which
differs from their \emph{real-time ordering}. As the logical ordering
is inherently dynamic, depending on properties such as
\jyscan\ missing a \jywrite, we formalize it in Hoare logic, by
keeping it as a list of events in auxiliary state that can be
dynamically reordered as needed. For example, Figure~\ref{fig:reorder}
shows the situation in the execution of \jyscan~that we reviewed
above. We start with the (initializing) writes of $5$ and $0$ already
executed, and our program performs the writes of $2$, $3$ and $1$ in
the real time order shown by the position of the events on the dashed
lines. In Figure~\ref{fig:reorder:before}, the logical order
$\ordlist$ coincides with real-time order, but is unsound for the
snapshot $\x=2, \y=1$ that \jyscan~wants to return. In that case, the
auxiliary code with which we annotate \jyscan, will change the
sequence $\ordlist$ in-place, as shown in
Figure~\ref{fig:reorder:after}.

Our specification and verification challenge then lies in reconciling
the following requirements. First, we have to posit specs that
say that \jywrite\ performs a write, and \jyscan\ performs a scan of
the memory, with the operations executing in a single logical
moment. Second, we need to implement the event reordering discipline
so that a method call only reorders events that overlap with it; the
logical order of the past events should be preserved. This will be
accomplished by introducing yet further structures into the auxiliary
state and code. Finally, the specs must hide the specifics of
the reordering discipline, which should be internal to the snapshot
object. Different snapshot implementations should be free to implement
different reorderings, without changing the method specs.


\section{Specification}
\label{sc:formal}


\def\histx{\hist_\x}
\def\histy{\hist_\y}
\def\histp{\hist_p}

\newcommand{\sx}{S_\x}
\newcommand{\sy}{S_\y}
\newcommand{\spp}{S_p}
\newcommand{\sss}{S_s}
\newcommand{\wx}{W_\x}
\newcommand{\wy}{W_\y}
\newcommand{\wpp}{W_p}


\def\toff{t_{\mathsf{off}}}

\subparagraph*{General considerations.}
For the purposes of specification and proof, we record a history of
the snapshot object as a set of entries of the form $t \mapsto (p,
v)$. The entry says that at time $t$ (a natural number), the value $v$
was written into the pointer $p$. We thus identify a write event with
a \emph{single} moment in time $t$, enabling the specs of
\jywrite\ and \jyscan\ to present the view that write events are
logically atomic.
Moreover, in the case of snapshots, we can ignore the scan events in
the histories. The latter do not modify the state in a way observable
by clients who can access the shared pointers only via interface
methods \jywrite\ and\ \jyscan.

We keep three auxiliary history variables. The history variables
$\histS$ and $\histO$ are local to the specified thread, and record
the \emph{terminated} write events carried out by the specified
thread, and that thread's interfering environment, respectively. We
refer to $\histS$ as the \emph{self}-history, and to $\histO$ as the \emph{other}-history~\cite{LeyWildN+POPL13,NanevskiLSD+ESOP14,OPLSS:Notes,SergeyNB+ESOP15}. The
role of $\histO$ is to enable the spec of \jywrite\ to situate the
performed write event within the larger context of past and ongoing
writes, and the spec of \jyscan\ to describe how it logically
reordered the writes that overlapped with it.
The third history variable $\histJ$ records the set of write events
that are in progress. These are events that have been initiated,
timestamped, and have executed their physical write to memory, but
have not terminated yet. It is an important component of our auxiliary
state design that when a write event terminates, it is moved from
$\histJ$ to the invoking thread's $\histS$, to indicate the
\emph{ownership} of the write by the invoking thread.
We name by $\hist$ the union $\histS \hunion \histO \hunion \histJ$,
which is the global history of the data structure. As common in
separation logic, the union is \emph{disjoint}, \ie, it is undefined
if the components contain duplicate timestamps. By the semantics of
our specs, $\hist$ is always defined, thus $\histS$, $\histO$ and
$\histJ$ never duplicate timestamps.

The real-time ordering of the timestamped events is the natural
numbers ordering on the timestamps. To track the \emph{logical}
ordering, we need further auxiliary notions.
The first is the auxiliary variable $\ordlist$, whose type is a
mathematical sequence. The sequence $\ordlist$ is a permutation of
timestamps from $\hist$ showing the logical ordering of the events in
$\hist$. We write $t_1 \tleq t_2$, and say that $t_1$ is logically
ordered before $t_2$, if $t_1$ appears before $t_2$ in $\ordlist$. The
sequence $\ordlist$ resides in joint state, and can be dynamically
modified by any thread. For example, the execution of the scanner may
reorder $\ordlist$, as shown in
Figure~\ref{fig:reorder:after}. Because $\ordlist$ is a sequence, the
order $\tleq$ is linear.

%
%

Because sequence $\ordlist$ changes dynamically under interference, it
is not appropriate for specifications. Thus, our second auxiliary
notion is the \emph{partial} order $\stableorder$, a suborder of
$\tleq$ that is \emph{stable} in the following sense. It relates the
timestamps of events whose logical order has been determined,
\emph{and will not change in the future}. Thus $\stableorder$ can grow
over time, to add new relations between previously unrelated
timestamps, but cannot change the old relations.

%

To illustrate the distinction between the two orders, we refer to
Figure~\ref{fig:reorder:before}. There, $\ordlist$ represents the
linear order $5{-}0{-}2{-}3{-}1$, which changes in
Figure~\ref{fig:reorder:after} to $5{-}0{-}2{-}1{-}3$.
Since $1$ and $3$ exchange places, the stable order $\stableorder$
cannot initially relate the two. Thus, in
Figure~\ref{fig:reorder:before}, $\stableorder$ is represented by the
Hasse diagram $5{-}0{-}2{<}\!\begin{array}[c]{c}1\\ 3\end{array}$. In
Figure~\ref{fig:reorder:after}, the relation $1{-}3$ is added to this
partial order, making it the linear order $5{-}0{-}2{-}1{-}3$. Note
how the previous relations remain unchanged.

\newcommand{\scanned}[1]{\mathsf{scanned}\,#1}

The third auxiliary notion is the set $\scanned\stableorder$ of
timestamps. A write's timestamp is placed in $\scanned\stableorder$,
if that write has been observed by some scanner; that is, the written
value is returned in some snapshot, or has been rewritten by another
value that is returned in some snapshot. To illustrate, in the above
example, $\{5, 0, 2\} \subseteq \scanned\stableorder$.  Intuitively,
we want to model that after a write has been observed, the ordering of
the events logically preceding the write must be stabilized, and
moreover, must be a sequence. Thus,
$\scanned\stableorder$ is a \emph{linearly ordered subset} of
$\stableorder$.\footnote{In terminology of linearizability, one may
  say that $\scanned\stableorder$ is the set of ``linearized''
  writes.}  The set $\scanned\stableorder$ can also be seen as
\emph{representing all the scans that have already been
  executed}. Such representation of scans allows us to avoid tracking
scan events directly in the history.

In the sequel, we concretize $\stableorder$ and
$\scanned\stableorder$ in terms of $\ordlist$ and other auxiliary
state. However, we keep the notions abstract in the method specs and
in client reasoning. This enables the use of different snapshot
algorithms, with the same specs, without invalidating the client
proofs. We also mention that $\ordlist$, $\stableorder$ and
$\scanned\stableorder$ can be encoded as user-level concepts in FCSL,
and require no new logic to be developed.

%

\subparagraph*{Snapshot specification.}
\def\chain{\mathsf{chain}}
\def\eval{\mathsf{eval}}
\def\static{\mathsf{static}}

\newcommand\myddarrow{\mathrel{\rotatebox[origin=c]{270}{$\twoheadrightarrow$}}}

\newcommand{\sideal}[2]{#1\,{\myddarrow}\,#2}
\newcommand{\ideal}[2]{#1\,{\downarrow}\,#2}

\begin{figure}[t]
\centering
\[
\begin{array}{l}
\mathtt{write}\ (p, v) : 
\begin{array}[t]{l}
\tsPre{\{\histS = \hempty\}}\
\tsPos{\{\exists t\ldot \histS' = t \mapsto (p, v) \wedge
    \dom {\histO} \cup \scanned\, \stableorder
       \subseteq \sideal{\stableorderP}{t}\}} @ C 
\end{array}\\[5pt]
\mathtt{scan} : 
\tsPre{\{\histS = \hempty\}}\ 
\!\!\begin{array}[t]{l}
\tsPos{\{r\ldot \exists t\ldot \histSP = \hempty\! \wedge\!
   r =\! \eval\ t\, {\stableorderP}\, {\histP} \wedge\!
  \dom{\hist}\! \subseteq \ideal{\stableorderP}{t}\! \wedge\!
  t \in \scanned{\stableorderP}\}} @ C 
\end{array}
\end{array}
\]
\caption{\label{fig:specs} Snapshot method specification. 
}
\end{figure}

Figure~\ref{fig:specs} presents our specs for \jyscan~and
\jywrite. These are partial correctness specs that describe how
the methods change the state from the precondition (first braces) to
the postcondition (second braces), possibly influencing the value $r$
that the procedure returns. We use VDM-style notation with unprimed
variables for the state before, and primed variables for the state
after the method executes. We use Greek letters for state-dependent
values that can be mutated by the method, and Latin letters for
immutable variables.
The component $C$ is a state transition system (STS) that describes the
state space of the algorithm, i.e, the invariants on the auxiliary and
real state, and the transitions, i.e., the allowed atomic mutations of
the state. For now, we keep $C$ abstract, but will define it in
Sections~\ref{sc:auxiliaries} and~\ref{sc:implementation}.
We denote by $\ideal{\stableorder}{t}$ the downward-closed set of
timestamps
$\ideal{\stableorder}{t} = \{ s \mid s\ {\stableorder}\ t \}$. Let
$\sideal{\stableorder}{t} = (\ideal{\stableorder}{t})\setminus\{t\}$.


The spec for \jywrite\ says the following. The precondition starts
with the empty self history $\histS$, indicating that the procedure
has not made any writes. In the postcondition, a new write event $t
\mapsto (p, v)$ has been placed into $\histS'$. Thus, a call to
\jywrite\ wrote $v$ into pointer $p$. The timestamp $t$
is fresh, because $\histP$ does not contain duplicate timestamps.
%
%
Moreover, the write appears as if it occurred atomically at time
$t$, thus capturing the logical atomicity of \jywrite.

The next conjunct, $\dom{\histO} \cup \scanned\stableorder \subseteq
\sideal{\stableorderP}{t}$, positions the write $t$ into the context
of other events. In particular, if $s \in \dom{\histO}$, \ie, if $s$
finished prior to invoking \jywrite, then $s$ is logically ordered
strictly before $t$. In other words, \jywrite\ cannot reorder prior
events that did not overlap with it. The definition of linearizability
contains a similar prohibition on reordering non-overlapping events,
but here, we capture it using a Hoare-style spec. For similar reasons,
we require that $\scanned\stableorder \subseteq
\sideal{\stableorderP}{t}$. As mentioned before,
$\scanned\stableorder$ represents all the scans that finished prior to
the call to \jywrite. Consequently, they do not overlap with
\jywrite\ in real time, and have to be logically ordered before $t$.

Notice what the spec of \jywrite\ \emph{does not prevent}. It is
possible that some event, say with a timestamp $s$, finishes in real
time before the call of \jywrite~at time $t$. Events $s$ and $t$ do
not overlap, and hence cannot be reordered; thus
$s\ {\stableorder}\ t$ always. However, the relationship of $s$ with
other events that ran concurrently with $s$, may be fixed only later,
thus supporting implementation of ``future-dependent'' nature, such as
Jayanti's.



In the case of \jyscan, we start and terminate with an empty $\histS$,
because \jyscan\ does not create any write events, and we do not track
scan events. However, when \jyscan\ returns the pair $r = (r_\x,
r_\y)$, we know that there exists a timestamp $t$ that describes when
the scan took place. This $t$ is the timestamp of the last write
preceding the call to \jyscan.

The postcondition says that $t$ is the moment in which the snapshot
was logically taken, by the conjunct $r =
\eval\ t\ \stableorderP\ \histP$.  Here, $\eval$ is a pure,
specification-level function that works as follows. First, it reorders
the entire real-time post-history $\histP$ according to logical
post-ordering $\stableorderP$. Then, it computes and returns the
values of $x$ and $y$ that would result from executing the write
events of such reordered history up to the timestamp $t$. For example,
if $t$ is the timestamp of event $1$ in
Figure~\ref{fig:reorder:after}, then $\eval\ t\ \stableorderP\ \histP$
would return $(2, 1)$. Hence, the conjunct says that
\jyscan\ performed a scan of $x$ and $y$, consistent with the ordering
$\stableorderP$, and returned the read values into $r$. The scan
appears as if it occurred atomically, immediately after time $t$, thus
capturing the atomicity of \jyscan.


The next conjunct, $\dom{\hist} \subseteq \ideal{\stableorderP}{t}$,
says that the scanner returned a snapshot that is current, rather than
corresponding to an outdated scan. For example, referring to
Figure~\ref{fig:reorder}, if \jyscan\ is invoked after the events $2$
and $1$ have already executed, then \jyscan\ should not return the
pair $(5, 0)$ and have $t$ be the timestamp of the event $0$, because
that snapshot is outdated. Specifically, the conjunct says that the
write events from $\hist$ are ordered no later than $t$, similar to
the postcondition of \jywrite. However, while in \jywrite\ we
constrained the events from $\dom{\histO} \cup \scanned\stableorder$,
here we constrain the full global history $\hist = \histO \hunion
\histJ$. The addition of $\histJ$ shows that the scanner will observe
and order all of the write events that have been timestamped and
recorded in $\histJ$ (and thus, that have written their value to
memory), prior to the invocation of \jyscan.

Lastly, the conjunct $t \in \scanned\stableorderP$ explicitly says
that $t$ has been observed by the just finished call to scan.

Again, it is important what the spec does not prevent. It is possible
that the timestamp $t$ identified as the moment of the scan,
corresponds to a write that has been initiated, but has not yet terminated.
Despite being ongoing, $t$ is placed into $\scanned\stableorderP$
(\ie, $t$ is ``linearized''). Also, notice that the postcondition of
\jyscan\ actually specifies the ``linearization'' order of events that
are initiated by another method, namely \jywrite, thus supporting
implementations of ``non-local'' nature, such as Jayanti's.

We close the section with a brief discussion of how the specs are
used. Because $C$, $\stableorder$ and $\scanned$ are abstracted from
the clients, we need to provide an interface to work with them. The
interface consists of a number of properties showing how various
assertions interact, summarized in the statements below.

The first statement presents the invariants on the transitions of STS
$C$, often referred to as 2-state invariants.
Another way of working with such invariants is to include them in the
postcondition of every method.\footnote{In fact, this is what we
  currently do in our Coq files.} For simplicity, here we agglomerate
the properties, and use them implicitly in proofs as needed.
\begin{invariant}[Transition invariants]\label{inv:mono}
In any program respecting the transitions of $C$: 
\begin{enumerate}
\item\label{inv:mono:hist} $\hist \subseteq \histP$, $\histS \subseteq
  \histSP$, and $\histO \subseteq \histOP$.

\item \label{inv:mono:stable} $\stableorder \subseteq \stableorderP$
  and $\scanned{\stableorder} \subseteq \scanned{\stableorderP}$.

\item\label{inv:mono:ideal} For every $s \in \scanned{\stableorder}$,
  $\ideal{\stableorder}{s} = \ideal{\stableorderP}{s}$.
\end{enumerate}
\end{invariant}

Invariant~\ref{inv:mono}.\ref{inv:mono:hist} says that histories only
grow, but does not insist that $\histJ \subseteq \histJP$, as
timestamps can be removed from $\histJ$ and transferred to $\histS$.
Invariant~\ref{inv:mono}.\ref{inv:mono:stable} states that
$\stableorder$ is monotonic, and the same applies for
$\scanned{\stableorder}$. This is a fundamental stability requirement
for our system: no transition in the STS $C$ can change the relations
between write events in $\stableorder$ and, moreover, write events
which have been observed by the scanner--- and thus are in
$\scanned{\stableorder}$--- cannot be unobserved.
Invariant~\ref{inv:mono}.\ref{inv:mono:ideal} says that if a new event
is added to increase $\stableorder$ to $\stableorderP$, that event
appears logically later than any $s \in \scanned{\stableorder}$. In
other words, once events are observed by a scanner, and placed into
$\scanned{\stableorder}$ in a certain order, we cannot insert new
events among them to modify the past observation.

The second statement exposes the properties of $\stableorder$,
$\scanned$, and $\mathsf{eval}$ that are used for client reasoning:
\begin{invariant}[Relating $\scanned$ and snapshots]\label{lem:scanned}
The set $\scanned\, \stableorder$ satisfies the following properties:
\begin{enumerate}
 \item\label{lem:scanned:total} if $ t_1 \in
   \scanned{\stableorder}$ and $t_2 \in \scanned{\stableorder}$, then
   $ t_1 \mathrel{\stableorder} t_2 \vee t_2 \mathrel{\stableorder}
   t_1 $ (linearity).
  
 \item \label{lem:scanned:wkn} if $ t_2 \in
   \scanned{\stableorder}$ and $ t_1 \mathrel{\stableorder} t_2$, then
   $t_1 \in \scanned{\stableorder}$ (downward closure).

   
\item \label{lem:scanned:eval} if $t \in \scanned{\stableorder}$,
  $\hist \subseteq \histP$, $\stableorder \subseteq \stableorderP$,
  $\scanned{\stableorder} \subseteq \scanned{\stableorderP}$, and $
  \ideal{\stableorder}{t} = \ideal{\stableorderP}{t}$ then $ \eval\,
  t\, \stableorder\, \hist = \eval\, t\, \stableorderP\,
  \histP$. (snapshot preservation).
\end{enumerate}
\end{invariant}

The first two properties merely state that the subset
$\scanned{\stableorder}$ is totally-ordered
(\ref{lem:scanned}.\ref{lem:scanned:total}) and also downward closed
(\ref{lem:scanned}.\ref{lem:scanned:total}). The last property is the
most interesting: it entails that once a snapshot is observed by
\jyscan, its validity will not be compromised by future or ongoing
calls to \jywrite. Thus, snapshots returned from previous calls to
\jyscan~are still valid and observable in the future.


\section{Client reasoning}
\label{sc:clients}


\subparagraph*{Comparison with linearizability specifications.}
In linearizability one would specify \jywrite\ and \jyscan\ by
relating them, via a simulation argument, to sequential programs for
writing and scanning, respectively. On the face of it, such specs are
indeed simpler than ours above, as they merely state that \jywrite\
writes and \jyscan\ scans. Our specs capture this property with one
conjunct in each postcondition. The remainders of the postconditions
describe the relative order of the atomic events, \emph{observed} by
threads, including explicit prohibition on reordering non-overlapping
events, which is itself inherent in the definition of linearizability.

However, the additional specifications are not pointless, and they
become useful when it comes to reasoning about
clients. Linearizability tells us that we can simplify a fine-grained
client program by replacing the occurrences of \jywrite\ and
\jyscan\ with the atomic and sequential equivalents, thus turning the
client into an equivalent coarse-grained concurrent program. However,
linearizability is not directly concerned with verifying that
coarse-grained equivalent itself.
%
%
%
Then, if one is interested in proving client properties which involve
timing and/or ordering properties of such events, it is likely that
the simple sequential spec described above do not suffice, and extra
auxiliary state is still required.

On the other hand, if one wants to reason about such clients using a
Hoare logic, then our specs are immediately useful. Moreover, in our
setting, client reasoning depends solely on the API for scan and
write, regardless of the different linearizations of a program. In the
sequel, we illustrate this claim by deriving interesting client timing
properties out of the specs of \jywrite\ and \jyscan.


Moreover, because we use separation logic, our approach easily
supports reasoning about programs with a dynamic number of threads,
and about programs that transfer state ownership. In fact, as we
already commented in Section~\ref{sc:formal}, our proofs rely on
transferring write events from $\histJ$ (joint ownership) to $\histS$
(private ownership), upon \jywrite's termination.
This is immediate in FCSL, as reasoning about histories inherits the
infrastructure of the ordinary heap-based separation logic, such as
framing and, in this case, ownership transfer.
In contrast, Linearizability is usually considered for a fixed number
of threads, and its relationship with ownership transfer is more
subtle~\cite{GotsmanY12+CONCUR12, CeroneGY+ICALP14}.

An additional benefit of specifying the event orders by Hoare triples
at the user level, is that one can freely combine methods with
different event-ordering properties, that need not respect the
constraints of linearizability~\cite{SergeyNBD+OOPSLA16}.





\subparagraph*{Example clients.}

We first consider the client $e$, defined as follows:
\[
\hfill\begin{array}{l||l||l}
        \begin{array}{c}
         \esc{write}~(x,2);\\ 
         \esc{write}~(y,1)~ 
        \end{array}~
& ~\esc{scan}~()~ &~\esc{write}~(x,3)  
\end{array}\hfill
\]
It is our running example from Figure~\ref{fig:weird:code}. We will
show that it satisfies the spec below. In the sequel we omit the STS
$C$, as it never changes.

\[
\!\!\!\begin{array}{l}
e : 
  \tsPre{\{ \histS = \hempty \}}
\!\!\begin{array}[t]{l}
  \tsPos{\{ r\ldot \exists t_1\, t_2\, t_3\, t_s \ldot\,
    \histS' = t_1 \hpts (y,1) \hunion t_2 
    \hpts (x,2) \hunion t_3 \hpts (x,3)\wedge
    \dom{\hist} \subseteq \ideal{\stableorderP}{t_s}\wedge\!}\\
  \tsPos{\hphantom{\{ r\ldot \exists t_1\, t_2\, t_3\, t_s \ldot}\, 
    \dom{\histO} \subseteq \sideal{\stableorderP}{t_2},
    \sideal{\stableorderP}{t_3} \wedge 
%
    t_2\ {\stableorderP}\ t_1 \wedge r = \eval\ t_s\ \stableorderP\ \histP \}}\\
\end{array}
\end{array}
\]

The spec of $e$ states that (1) $\esc{write}~(x,2)$, timestamped
$t_2$, occurs sequentially before $\esc{write}~(y,1)$ which is
timestamped $t_1$, (2) the remaining write, timestamped $t_3$, and the
scan, timestamped $t_s$, are not temporally constrained, and (3) the
writes that terminated before the client started are ordered before
$t_2$ (and thus before $t_1$), $t_3$ and $t_s$. The example
illustrates how to track timestamps and their order, but does not
utilize the properties of $\scanned{\stableorder}$. We illustrate the
latter in another example at the end of this section.

We first verify the subprograms $\esc{scan}~()~\parallel
\esc{write}~(x,3)$ and $\esc{write}~(x,2);\ \esc{write}~(y,1)$
separately, and then combine them into the full proof. As proof
outlines show intermediate, in addition to pre- and post-state, we
cannot quite utilize VDM notation in them. As a workaround, we
explicitly introduce logical variables $h$ and $h_o$ to name (subsets
of) the initial global and other history.
{
\[
\!\!\!\!\!\begin{array}{l c} \num{1} &
 {\specK{\{ \histS = \hempty \wedge h \subseteq \hist \wedge h_o
                           \subseteq \histO \}}} \\[3pt]
& \!\!\!\!\!\! \begin{tabular}{l || l}
     \begin{array}[t]{l l}
       \num{2a} &
       \specK{\{ \histS = \hempty \wedge h \subseteq \hist \wedge h_o \subseteq \histO \}}\\
       \num{3a} & \multicolumn{1}{c}{\esc{scan}~()} \\
       \num{4a} &
       \specK{\{ r\ldot\, \exists\, t_s\ldot
         \histS = \hempty \wedge  \dom{h} \subseteq \ideal{\stableorder}{t_s} \wedge \hbox{}}\\
       &
       \specK{\phantom{\{r\ldot\, \exists\, t_s\ldot}\,
         r = \eval\, t_s\, \stableorder\, \hist\}}
     \end{array}~&~
     \!\!\!\begin{array}[t]{r l}
       \num{2b} &
       \specK{\{ \histS = \hempty \wedge h \subseteq \hist \wedge h_o \subseteq \histO \}} \\       
       \num{3b} & \multicolumn{1}{c}{\esc{write}~(x,3)}\\
       \num{4b} &
       \specK{\{ \exists\, t_3\ldot
         \histS = t_3 \hpts (x,3) \wedge \hbox{}}\\
       & \specK{\phantom{\{\exists\, t_3\ldot}\,
         \dom{h_o} \subseteq \sideal{\stableorder}{t_3}\}}
     \end{array}
   \end{tabular}\\[3pt]
  \num{5} & \hfill {\specK{\{ r\ldot\, \exists t_3\, t_s\ldot\,
    \histS = t_3 \hpts (x,3) \wedge
    \dom{h_o} \subseteq \sideal{\stableorder}{t_3} \wedge 
    \dom{h} \subseteq \ideal{\stableorder}{t_s} \wedge
    r = \eval\ t_s\ \stableorder\ \hist\}}} \hfill
\end{array}
\]}

The proof applies the rule for parallel composition of FCSL. This rule
is described in Appendix~\ref{sc:background}.
Here, we just mention that, upon forking, the rule distributes the
value of $\histS$ of the parent thread, to the $\histS$ values of its
children; in this case, all these are $\hempty$. Dually, upon joining,
the $\histS$ values of the children in lines 4a and 4b, are collected,
in line 5, into that of the parent. The other assertions in 4a and 4b
directly follow from the specs of \jyscan\ and \jywrite\ and the
Invariants~~\ref{inv:mono}.\ref{inv:mono:hist}
and~~\ref{inv:mono}.\ref{inv:mono:stable}, and directly transfer to
line 5.
While the proof outline does not establish how
\jyscan\ and \jywrite\ interleaved, it establishes that $t_3$ and $t_s$
both appear after the writes that are prior to the client's call.
{
\[
\begin{array}{r l}
  \num{1} & \specK{\{ \histS = \hempty \wedge h \subseteq \hist \wedge
    h_o \subseteq \histO \}} \\ 
  \num{2} & \esc{write}~(x,2);\\
  \num{3} & \specK{\{ \exists\, t_2 \ldot\, \histS = t_2 \hpts (x,2) \wedge \dom{h_{\othersub}} \subseteq \sideal{\stableorder}{t_2} \}} \\
  \num{4} & \esc{write}~(y,1)\\
  \num{5} & \specK{\{ \exists\, t_1\, t_2\ldot\, 
    \histS = t_1 \hpts (y,1) \hunion t_2 \hpts (x,2) \wedge 
    \dom{h_{\othersub}} \subseteq \sideal{\stableorder}{t_2} \wedge 
    t_2\ {\stableorder}\ t_1\}}
\end{array}
\]}

The second proof outline starts with the same precondition. Then line
$3$ directly follows from the spec of \jywrite, using $h_o
\subseteq \histO$. To proceed, we need to apply FCSL
\emph{framing}: the precondition of \jywrite\ requires $\histS =
\emptyset$, but we have $\histS = t_2 \mapsto (x, 2)$.
The frame rule is explained in Appendix~\ref{sc:background}.
Here we just mention that framing modifies the spec of \jywrite\ by
joining $t_2 \mapsto (x, 2)$ to $\histS$, $\histSP$ \emph{and}
$\histO$ as follows.
\[
\mathtt{write}\ (p, v) : 
\begin{array}[t]{l}
\tsPre{\{\histS = t_2 \mapsto (x, 2)\}}\
\tsPos{\{\exists t\ldot \histS' = t \mapsto (p, v) \hunion t_2 \mapsto (x, 2) \wedge \hbox{}}\\
\tsPos{\hphantom{\{\histS = t_2 \mapsto (x, 2)\}\ \{\exists t\ldot}
    \dom {\histO \hunion t_2 \mapsto (x, 2)} \cup \scanned\, \stableorder
       \subseteq \sideal{\stableorderP}{t}\}}
\end{array}
\]

Such a framed spec for \jywrite\ gives us that after line 4: (1)
$\histS = t_1 \hpts (y,1) \hunion t_2 \hpts (x,2)$, and (2)
$\dom{h_{\othersub} \hunion t_2 \mapsto (x, 2)} \subseteq
\sideal{\stableorder}{t_1}$. From Invariants~\ref{inv:mono}, we also
obtain that (3) $\dom{h_{\othersub}} \subseteq
\sideal{\stableorder}{t_2}$, which simply transfers from line 3. Now,
in the presence of (2), we can simplify (3) into
$t_2\ {\stableorder}\ t_1$, thus obtaining the postcondition in line
5.

The final step applies the rule for parallel composition to the two
derivations, splitting $\histS$ upon forking, and collecting it upon
joining:

\[
e : \begin{array}[t]{l}
    \specK{\{\histS = \hempty \wedge h \subseteq \hist \wedge h_o \subseteq \histO\}}\\
 \specK{\{r\ldot \exists\, t_1\, t_2\, t_3\, t_s \ldot 
    \histS = t_1 \hpts (y,1) \hunion t_2 \hpts (x,2) \hunion t_3 \hpts
        (x, 3) \wedge  \dom{h} \subseteq \ideal{\stableorder}{t_s} \wedge \hbox{}}\\
\specK{\hphantom{\{r\ldot \exists\, t_1\, t_2\, t_3\, t_s \ldot }
\dom{h_{\othersub}} \subseteq \sideal{\stableorder}{t_2}, \sideal{\stableorder}{t_3} \wedge 
 t_2\ {\stableorder}\ t_1 \wedge 
     r = \eval\, t_s\, \stableorder\, \hist\}}
\end{array}
\]

From here, the VDM spec of $e$ is derived by priming the Greek letters
in the postcondition, and choosing $h = \hist$ and $h_o = \histO$.

The spec of $e$ can be further used in various contexts. For example,
to recover the context from Section~\ref{sc:overview}, where $e$ is
invoked with $x = 5$, $y = 0$, we can frame $e$ wrt.~$\histS = t_5
\hpts (x, 5) \hunion t_0 \hpts (y, 0)$ to make explicit the events
that initialize $x$ and $y$. Then, 
%
%
it is possible to derive in FCSL that if $e$ executes without
interference (\ie, if $\hist = \histO = \histP = \histOP = \hempty$),
then the result at the end must be $r \in \{(5,0), (2,0), (3,0),
(2,1), (3,1)\}$. As expected, $r \neq (5, 1)$, because the write of
$2$ sequentially precedes the write of $1$.

We next illustrate the use of Invariants~\ref{lem:scanned}, which are
required for clients that use $\esc{scan}$ in \emph{sequential
  composition}. We consider the program
\[
\hfill{
e' = r \tbnd \esc{scan} ;\, \esc{write}\, (x,v);\, {\small{\kw{return}}}\, r 
}\hfill
\]
and prove that $e'$ can be ascribed the following spec:
\[
\hfill
e' : \tsPre{\{ \histS = \hempty \}} \
\begin{array}[t]{l}
     \tsPos{\{ \exists\, t_s\ t_x\ldot\,
       \histSP = t_x \hpts (x, v) \wedge
       t_s \in \sideal{\stableorderP}{t_x} \wedge 
           r = \eval\ t_s\ \stableorderP\ \histP \}}
\end{array}\hfill
\]

The spec says that the write event ($t_x$) is subsequent to the scan
($t_s$), as one would expect. In particular, the snapshot $r$ remains
valid, \ie, the write does not change the order $\stableorder$ and
history $\hist$ in a way that makes $r$ cease to be a valid snapshot
in $\stableorderP$ and $\histP$. The proof outline follows, with the
explanation of the critical steps.
\[
\!\begin{array}{r l}
 \num{1} & \specK{\{ \histS = \hempty \}}\\ 
 \num{2} & r \tbnd \esc{scan};\\
 \num{3} & \specK{\{ \exists\, t_s, w'(=\stableorder), h'(=\hist)\ldot 
       \histS = \hempty \wedge 
        t_s \in \scanned{w'}  \wedge r = \eval\ t_s\ w'\ h'\}}\\
 \num{4} & \esc{write}\, (x,v);\\
 \num{5} & \specK{\{ \exists\, t_s\, t_x\ldot\,
               \histS = t_x \hpts (x, v) \wedge t_s \in \sideal{\stableorder}{t_x} \wedge t_s \in \scanned{\stableorder} \wedge
               r = \eval\ t_s\ \stableorder\ \hist\}}\\
 \num{6} & \kw{return}\ r
\end{array}
\]

Line 3 is a direct consequence of the spec of \jyscan, where we
omitted the conjunct $\dom{\hist} \subseteq
\ideal{\stableorderP}{t_s}$, as we do not need it for the subsequent
derivation. We also introduce explicit names $w'$ and $h'$ for the
current values of $\stableorder$ and $\hist$.
Now, to derive line 5, by the spec of \jywrite, we know there exists a
timestamp $t_x$ corresponding to the write, such that (1) $\histS =
t_x \hpts (x, v)$, which is a conjunct in line 5, and also (2)
$\dom{\histO} \cup \scanned{w'} \subseteq
\sideal{\stableorder}{t_x}$. Furthermore, (3) $t_s \in \scanned{w'}$,
and (4) $r = \eval\ t_s\ w'\ h'$, simply transfer from line 3. From
(2) and (3), we infer that $t_s \in \sideal{\stableorder}{t_x}$. To
complete the derivation of line 5, it remains to show that $t_s \in
\scanned{\stableorder}$ and $r = \eval\ t_s\ \stableorder\ \hist$. For
this, we use (3), (4) and the Invariants~\ref{inv:mono}
and~\ref{lem:scanned}, as follows.  First, by
Invariant~\ref{inv:mono}.\ref{inv:mono:ideal}, and because $t_s \in
\scanned{w'}$, we get $\ideal{w'}{t_s} =
\ideal{\stableorder}{t_s}$. By
Invariant~\ref{inv:mono}.\ref{inv:mono:stable}, this gives us $t_s \in
\scanned{\stableorder}$ as well. By
Invariant~\ref{inv:mono}.\ref{inv:mono:hist}, $h' \subseteq \hist$,
and then by Invariant~\ref{lem:scanned}.\ref{lem:scanned:eval}, $r =
\eval\ t_s\ w'\ h' = \eval\ t_s\ \stableorder\ \hist$, completing the
deduction of line 5.


Observe that the main role of $\scanned$ in proofs is to enable
showing \emph{stability} of values obtained by $\eval$, using
Invariant~\ref{lem:scanned}.\ref{lem:scanned:eval}. The remaining
Invariants~\ref{lem:scanned}.\ref{lem:scanned:total}
and~\ref{lem:scanned}.\ref{lem:scanned:wkn} allow us to replace a
number of conjuncts about $\scanned$ by a single one that expresses
the membership of the largest timestamp in the current $\scanned$ set.

\section{Internal auxiliary state}
\label{sc:auxiliaries}

In order to verify the \emph{implementations} of \jywrite~and \jyscan,
we require further auxiliary state that does \emph{not} feature in the
specifications, and is thus hidden from the clients.

First, we track the point of execution in which \jywrite~and
\jyscan~are, but instead of line numbers, we use datatypes to encode
extra information in the constructors.
For example, the scanner's state is a triple $(\sss, \sx,
\sy)$. $\sss$ is drawn from $\{ \sOn, \sOff\, t\}$. If $\sOn$, then
the scanner is in lines~\lineScanSetsS--\lineScanReadsY~in
Figure~\ref{fig:jayanti-snapshot}. If $\sOff\ t$, the the scanner
reached line~\lineScanUnsetsS~at ``time'' $t$, and is now in
\lineScanReadsFX--\lineScanRelinks. $\sx$ is a boolean bit, set when
the scanner clears $\fx$ in line~\lineScanClearsX, and reset upon
scanner's termination (dually for $\sy$ and $\fy$).
Writers' state for $x$ is tracked by the auxiliary $\wx$ (dually,
$\wy$). These are drawn from $\{\wInit, \wWrite\ t\, v, \wDirty\ t\,
v, \wClean\ t\, v \}$, where $t$ marks the beginning of the write and
$v$ is the value written to pointer $p$. If $\wInit$, then no write is
in progress. If $\wWrite\ t\, v$, then the writer is in
line~\lineWrtWrt. If $\wDirty\ t\, v$, then $b$ has been set in
line~\lineWrtChk, triggering forwarding. If $\wClean\ t\, v$, the
writer is free to exit.



Second, like in linearizability, we record the ending times of
terminated events, using an auxiliary variable $\E$. $\E$ is a
function that takes a timestamp identifying the beginning of some
event, and returns the ending time of that event, and is undefined if
the event has not terminated. However, we do not \emph{generate} fresh
timestamps to mark event ending times. Instead, at the end of
\jywrite, we simply read off the last used timestamp in $\hist$, and
use it as the ending time of \jywrite. This is a somewhat non-standard
way of keeping time, but it suffices to prove that events $t_1$ and
$t_2$ which are non-overlapping (\ie, $\E(t_1) < t_2$ or $\E(t_2) <
t_1$) are never reordered. The latter is required by the
postconditions of \jywrite\ and \jyscan, as we discussed in
Section~\ref{sc:formal}. 
%
%
%
Formally, the following is an \emph{invariant} of the snapshot object;
\ie, a property of the state space of STS $C$ from
Figure~\ref{fig:specs}, preserved by $C$'s transition.

\begin{invariant}\label{inv:overlap}%
The logical order $\tle$ preserves the real time order of
non-over\-lap\-ping events: $\forall t_1 \in \dom{\E}, t_2 \in
\dom{\hist}$, if $\E(t_1) < t_2$ then $t_1 \tle t_2$.
\end{invariant}

Third, we track the rearrangement status of write events wrt.~an
ongoing \emph{active} scan, by \emph{colors}. A scan is \emph{active}
if it has cleared the forwarding pointers in
lines~\lineScanClearsX\ and\ \lineScanClearsY, and is ready to read
$x$ and $y$. We keep the auxiliary variable $\C$, which is a function
mapping each timestamp in $\hist$ to a color, as follows.
%
%
\begin{itemize}
\item {\sf Green} timestamps identify write events whose position in
  the logical order is fixed in the following sense: if $\C(t_1) =
  \mathsf{green}$ and $t_1 \tle t_2$, then $t_1 <_{\ordlist'} t_2$ for
  every $\ordlist'$ to which $\ordlist$ may step by auxiliary code
  execution (Section~\ref{sc:implementation}). For example, since we
  only reorder overlapping events, and only the scanner reorders
  events, every event that finished before the active scan started
  will be green. Also, a green timestamp never changes its color.

\item {\sf Red} timestamps identify events whose order is not fixed,
  but which will \emph{not} be manipulated by the active scan, and are
  left for the next scan.

\item {\sf Yellow} timestamps identify events whose order is not fixed
  yet, but which \emph{may} be manipulated by the ongoing active scan,
  as follows.  The scan can \emph{push} a yellow timestamp in logical
  time, \emph{past} another green or yellow timestamp, but not past a
  red one. \emph{This is the only way the logical ordering can be
    modified.}

%
\end{itemize}

There are a number of invariants that relate colors and timestamps. We
next list the ones that are most important for understanding our
proof. We use $\histp$ to denote the sequence of writes into the
pointer $p$ that appear in the history $\hist$, sorted by their order
in $\ordlist$\footnotemark.

\footnotetext{For reasoning purposes, it serves us better to think of
  $\histp$ as sub-histories, with an {\emph external} ordering given
  by $\ordlist$. We do, however, implement $\histp$ as a list filter:
  $\histp =\, {\sf filter}\ (\lambda\, t\ldot t \hpts (p, \_) \in
  \hist)\ \ordlist$.}


\begin{invariant}[Colors]\label{inv:color}%
The colors of $\histp$ are described by the regular expression $\GYR$:
there is a non-empty prefix of green timestamps, followed by \emph{at
  most} one yellow, and arbitrary number of reds.
\end{invariant}

By the above invariant, the yellow color identifies the write event
into the pointer $p$, that is the \emph{unique} candidate for
reordering by the ongoing active scan. Moreover, all the writes into
$p$ prior to the yellow write, will have already been colored green
(and thus, fixed in time), whether they overlapped with the scanner or
not.

\begin{invariant}[Color of forwarded values]\label{inv:readFP}%
Let $\sss= \sOff\ \toff$, and $p \in \{x, y\}$, and $\spp = \TT$ (\ie,
scanner is in lines~\lineScanReadsFX--\lineScanChoosesRY), and $v \neq
\bot$ has been forwarded to $p$; \ie, $\aleksfwdp{p} \hpts v$. Then
the event of writing $v$ into $p$ is in the history, \ie, there exists
$t$ such that $t \hpts (p, v) \in \histp$. Moreover, $t$ is the last
green, or the yellow timestamp in $\histp$.
\end{invariant}

The above invariant restricts the set of events that could have
forwarded a value to the scanner, to only two: the event with the
(unique) yellow timestamp, or the one corresponding to the last green
timestamp. By Invariant~\ref{inv:color}, these two timestamps are
consecutive in $\histp$.


\begin{invariant}[Red zone]\label{inv:redzone}%
If $\sss = \sOff\ \toff, \sx = \TT, \sy = \TT$, then $\hist$ satisfies
the $\RZ$ pattern. Moreover, for every $t \in \dom{\hist}$:
\begin{itemize}
\item $\C(t) = \mathsf{green} \implies t \leq \toff$
\item $\C(t) = \mathsf{yellow} \implies t \leq \toff \leq \E (t)$
\item $\C(t) = \mathsf{red} \implies \toff < t$  
\end{itemize}
\end{invariant}

This invariant restricts the global history $\hist$ (not the
pointer-wise projections $\histp$). First, the red events in $\hist$
are consecutive, and cannot be interspersed among green and yellow
events. Thus, when a scanner pushes a yellow event past a green event,
or past another yellow event, it will not ``jump over'' any
reds. Second, the invariant relates the colors to the time $\toff$ at
which the scanner was turned off (in line~\lineScanUnsetsS,
Figure~\ref{fig:jayanti-snapshot}). This moment is important for the
algorithm; \eg, it is the linearization point for \jyscan~in
Jayanti's proof~\cite{Jayanti+STOC05}.
We will use the above inequalities wrt.~$\toff$ in our proofs, to
establish that the events reordered by the scanner \emph{do} overlap,
as per Invariant~\ref{inv:overlap}.

We can now define the stable logical order $\stableorder$, and the set
$\scanned{\stableorder}$, using the internal auxiliary state of colors
and ending times.
\begin{definition}[Logical order $\stableorder$ and
    $\scanned{\stableorder}$]\label{def-jleq}
  
\begin{enumerate}
\item $t_1\,{\stableorder}\,t_2 \eqdef (t_1 = t_2) \vee (\E(t_1) <
  t_2) \vee (t_1 \tle t_2 \wedge \C(t_1) = \mathsf{green})$
\item  $\scanned{\stableorder} = \{t \mid \ideal{\stableorder}{t} = \ideal{{\tleq}}{t} \wedge \forall s\in \ideal{\stableorder}{t}\ldot \C(s) = \mathsf{green}\}$.
\end{enumerate}
\end{definition}

From the definition of $\stableorder$, notice that
$t_1\,{\stableorder}\,t_2$ is stable (\ie, invariant under
interference), since threads do not change the ending times $\E$, the
color of green events, or the order of green events in $\tle$, as we
already discussed. 
From the definition of $\scanned{\stableorder}$, notice that for every
$t \in \scanned{\stableorder}$, it must be that
$\ideal{\stableorder}{t}$ is a linearly-ordered set
wrt.~$\stableorder$, because it equals a prefix of the \emph{sequence}
$\ordlist$.

We close this section with a few technical invariants that we use in
the sequel.


\begin{invariant}[Last write]\label{inv:last-key}%
Let pointer $p \in \{\x,\, y\}$, and $\mathsf{last}_\ordlist\ \histp$
be the timestamp in $\histp$ that is largest wrt.~the logical order
$\tleq$. Then the contents of $p$ equals the value written by the
event associated with $\mathsf{last}_\ordlist\ \histp$. That is, $p
\mapsto \histp (\mathsf{last}_\ordlist \histp)$.
\end{invariant}


\begin{invariant}[Joint history]\label{inv:joint-hist}%
Let pointer $p \in \{\x,\, y\}$. If the writer for $p$ is active \ie
$\wpp \neq \wInit$, then the write event that it is performing is
timestamped and placed into joint history $\histJ$. Dually, if $t \in
\dom{\histJ}$, then the event $t$ is performed by the active writer
for $p$:
$$t \hpts (p,v) \in \histJ \iff \wpp = \wWrite\ t\, v \vee
\wpp = \wDirty\ t\, v \vee \wpp = \wClean\ t\, v $$
\end{invariant}


\begin{invariant}[Terminated events]\label{inv:dom-tau}%
Histories $\histO$ and $\histS$ store only terminated events, \ie,
events whose ending times are recorded in $\E$. Moreover, the codomain
of $\E$ is bounded by the maximal timestamp, in real time, in
$\dom{\hist}$:
\begin{enumerate}
\item $\dom{\E} = \dom{\histS} \hunion \dom{\histO}$.
\item $\forall a \in \dom{\E}\ldot\ \E(a) \leq
  \mathsf{max}\ (\dom{\hist})$.
\end{enumerate}
\end{invariant}



\begin{lemma}[Green/yellow read values]\label{lemma:first-read}%
Let $p \in \{x, y\}$. If the scanner state is $\sss= \sOn, \spp =
\TT$, \ie, the scanner is between
lines~\lineScanReadsX--\lineScanReadsY\ in
Figure~\ref{fig:jayanti-snapshot}, and $ p \hpts v$ in the physical
heap, then exists $t$ such that $ t \hpts (p, v) \in
\histp$. Moreover, $t$
is the last green or the yellow timestamp in $\histp$.
\end{lemma}

\begin{lemma}[Chain]\label{lemma:complete-green}%
If $t\,{\in}\,\dom{\hist}$ and $\C(\ideal{{\tleq}}{t})\,{=}\,\mathsf{green}$, 
then
$\ideal{\stableorder}{t}\ =\ \ideal{{\tleq}}{t}$.
\end{lemma}

\begin{figure}[t]
\centering
\begin{tabular}{c@{\hfill}c}
\begin{minipage}[t]{.5\textwidth}
\[
\begin{array}{ll}
\num{1}~ & ~~ \esc{write}\ (p,\, v)\ \{\\ 
\num{2}~ & ~~~~~ \lat\,\actwrite{p}{v}; \aux{register}(p,v)\rat;\\
\num{3}~ & ~~~~~ \lat\, b \tbnd \act{read}(S);\ \aux{check}(p,b) \rat;\\
\num{4}~ & ~~~~~ \kw{if}\ b\\
\num{5}~ & ~~~~~ \kw{then}\ \lat\,%
                           \actwrite{\aleksfwdp{p}}{v};\ \aux{forward}(p)\rat; \\
\num{5'}~           & ~~~~ \lat\,\aux{finalize}(p)\rat\}
\end{array}
\]
\end{minipage}
&
\begin{minipage}[t]{.5\textwidth}
\[
\begin{array}{rl}
\num{6}~  & \esc{scan} () : ( A \times A )~ \{ \\ 
\num{7}~  & ~~~ \lat\,\actwrite{\s}{\esc{true}};\ \aux{set}(\esc{true}) \rat;\\  
\num{8}~  & ~~~ \lat\,\actwrite{\fx}{\bot};\ \aux{clear}(\x)\rat;\\
\num{9}~  & ~~~ \lat\,\actwrite{\fy}{\bot};\ \aux{clear}(\y)\rat;\\
\num{10}~ & ~~~ \var{vx} \tbnd \lat \act{read}(\x) \rat ; \\
\num{11}~ & ~~~ \var{vy} \tbnd \lat \act{read}(\y) \rat;  \\
\num{12}~ & ~~~ \lat\,\actwrite{\s}{\esc{false}};\ \aux{set}(\esc{false})\rat;\\
\num{13}~ & ~~~ \var{ox} \tbnd \lat \act{read}(\fx) \rat;\\
\num{14}~ & ~~~ \var{oy} \tbnd \lat \act{read}(\fy) \rat;\\
\num{15}~ & ~~~ \var{rx} \tbnd \kw{if}\ (\var{ox} \neq\bot)\
                \kw{then}\ \var{ox}\ \kw{else}\ \var{vx};\\
\num{16}~ & ~~~ \var{ry} \tbnd \kw{if}\ (\var{oy} \neq\bot)\
                \kw{then}\ \var{oy}\ \kw{else}\ \var{vy};\\
\num{17}~ & ~~~ \lat\,\aux{relink}(\var{rx}, \var{ry});\
                \kw{return}\ (\var{rx}, \var{ry})\,\rat\}
\end{array}
\]
\end{minipage}
\end{tabular}
\caption{Snapshot procedures annotated with auxiliary code.}
\label{fig:fcsl-snapshot}
\end{figure}

\section{Auxiliary code implementation}
\label{sc:implementation}


Figure~\ref{fig:fcsl-snapshot} annotates Jayanti's procedures with
auxiliary code (typed in \emph{italic}), with $\langle\mbox{angle
braces}\rangle$ denoting that the enclosed real and auxiliary code
execute \emph{simultaneously} (\ie, atomically). The auxiliary code
builds the histories, evolves the sequence $\ordlist$, and updates the
color of various write events, while respecting the invariants from
Section~\ref{sc:formal}. Thus, it is the \emph{constructive} component
of our proofs. Each atomic command in Figure~\ref{fig:fcsl-snapshot}
represents one \emph{transition} of the STS $C$ from
Figure~\ref{fig:specs}.

%
%
%

The auxiliary code is divided into several procedures, all of which
are sequences of reads followed by updates to auxiliary variables. We
present them as Hoare triples in Figure~\ref{fig:auxcode}, with the
unmentioned state considered unchanged. The bracketed variables
preceding the triples (\eg, $[t, v]$) are logical variables used to
show how the pre-state value of some auxiliary changes in the
post-state. To symbolize that these triples \emph{define} an atomic
command, rather than merely stating the command's properties, we
enclose the pre- and postcondition in angle brackets $\langle
- \rangle$.



{
\begin{figure}[t]
\centering
\small
\[
\begin{array}{l@{\, :\ } l}
 \aux{register}(p,v) & 
  \begin{array}[t]{l}
   \langle \wpp = \wInit\rangle\\ 
   \langle\ordlistP = \mathsf{snoc}\ {\ordlist}\ t,\
     \histJP = \histJ \hunion t \hpts (p,v),\ \wppP = \wWrite\ t\, v,\\
   \phantom{\langle} \C' = \textrm{if}\ (\sss = \sOn) \& \spp\
                    \textrm{then}\ \C[t \mapsto \mathsf{yellow}]\
                    \textrm{else}\ \C[t \mapsto \mathsf{red}]\rangle \\
   \ \mbox{\small{where $t = \mathsf{fresh}\ \hist = \mathsf{max}\, (\dom{\hist})+1$}}
  \end{array} \\[2.5pt]
  \aux{check}\,(p,b) & [t, v]\ldot
  \!\begin{array}[t]{l}
  \langle\wpp = \wWrite\ t\, v\rangle\ 
  \langle\wppP = \textrm{if}\ b\
  \textrm{then}\ \wDirty\ t\, \,v\ \textrm{else}\ \wClean\ t\, v\rangle
 \end{array}\\[2.5pt]
  \aux{forward}\,(p) & [t, v]\ldot
  \!\begin{array}[t]{l}
   \langle\wpp = \wDirty\ t\,v \rangle\\ 
   \langle\wppP = \wClean\ t\,v,\, %
   \C' = \textrm{if}\ (\sss=\sOn) \& \spp\ \textrm{then}\ \C[t \mapsto \mathsf{green}]\ \textrm{else}\ \C\rangle
  \end{array}\\[2.5 pt]
  \aux{finalize}(p) & [t, v]\ldot
  \!\begin{array}[t]{l}
  \langle\wpp = \wClean\ t\, v, %
  t \hpts (p, v) \in \histJ \rangle\\
  \langle\wppP = \wInit,\, \histSP = \histS \hunion t \hpts (p,v),\, %
  \histJP\! = \histJ\setminus\{t\},\,
  \E'\! = \E \hunion\! t \hpts \mathsf{max}\, (\dom{\hist}) \rangle
 \end{array}
\end{array}
\]
\hrulefill
%
\small
\[
\begin{array}{l@{\, :\ }l}
  \aux{set}(b) &
  \begin{array}[t]{l}
        \langle \sss = \textrm{if}\ b\ \textrm{then}\ \sOff\,(\_) \ 
        \textrm{else}\ \sOn,\ \sx = \neg\, b, \sy = \neg\, b \rangle\\
        \langle \sss' = \textrm{if}\ b\ \textrm{then}\ \sOn\ %
        \textrm{else}\ \sOff\,(\mathsf{last}\ \hist),\
         \sx' = \neg\, b, \sy' = \neg\, b \rangle
\end{array}\\[2.5pt] 
   \aux{clear}(p) &
  \begin{array}[t]{l}
   \langle\sss = \sOn,\ \spp = \FF \rangle\\
   \langle\sss' = \sOn,\ \spp' = \TT,\
     \CP = \C[\histp \hpts \mathsf{green}] \rangle
  \end{array}\\[2.5pt]
   \aux{relink}(r_x, r_y) & [t_x, t_y]\ldot
    \begin{array}[t]{l}
    \langle\sss = \sOff(\_), %
      t_x \hpts (x, r_x), t_y \hpts (y, r_y) \in \hist, \sx = \sy = \TT, \\
     \hphantom{\langle} \forall p \in \{x,y\}\ldot \lgVy\ p\, t_p \rangle\\
        \langle\sss' = \sss, \sx'= \sy'=\FF,%
        \CP = \C[t_x, t_y \hpts \mathsf{green}],\\
      \ \ordlist' = \textrm{if}\ (d = \mathsf{Yes}\ x\ s)\
                \textrm{then}\ \aux{push}\ s\ t_y\ \ordlist\\
                 \phantom{\ \ordlist =\ } \textrm{else\ if}\
                 (d = \mathsf{Yes}\ y\ s)\ \textrm{then}\
                 \aux{push}\ s\ t_x\ \ordlist\ \textrm{else}\ \ordlist\rangle\\
  \quad \mbox{where $d = \aux{inspect}\ t_x\, t_y\, \ordlist\, \C$}
  \end{array}
\end{array}
\]
\caption{\label{fig:auxcode} Auxiliary procedures for
\jywrite~and 
  \jyscan. Bracketed variables (\eg, $[t, v]$) are logical variables
  that scope over precondition and postcondition.}
\end{figure}
}

\subparagraph*{Auxiliary code for \jywrite.}
In line~\lineWrtWrt, $\aux{register}(p, v)$ creates the write event
for the assignment of $v$ to $p$. It allocates a \emph{fresh}
timestamp $t$, inserts the entry $t \mapsto (p, v)$ into $\histJ$, and
adds $t$ to the end of $\ordlist$, thus registering $t$ as the
currently latest write event. The fresh timestamp $t$ is computed out
of the history $\hist$; we take the largest natural number occurring
as a timestamp in $\hist$, and increment it by $1$.  The variable
$\wpp$ updates the writer's state to indicate that the writer finished
line~\lineWrtWrt\ with the timestamp $t$ allocated, and the value $v$
written into $p$. The color of $t$ is set to yellow (\ie, the order of
$t$ is left undetermined), but only if $(\sss= \sOn) \& \spp$ (\ie, an
active scanner is in line 10). Otherwise, $t$ is colored red,
indicating that the order of $t$ will be determined by a future scan.

In line~\lineWrtChk, $\aux{check}(p,b)$, depending on $b$, sets the
writer state to $\wDirty$, indicating that a scan is in progress, and
the writer should forward, or to $\wClean$, indicating that the writer
is ready to terminate.

In line~\lineWrtFwd, \aux{forward} colors the allocated timestamp $t$
green, if an active scanner has passed
lines~\lineScanClearsX--\lineScanClearsY~and is yet to reach
line~\lineScanUnsetsS, because such a scanner will definitely see the
write, either by reading the original value in
lines~\lineScanReadsX--\lineScanReadsY, or by reading the forwarded
value in lines~\lineScanReadsFX--\lineScanReadsFY. Thus, the logical
order of $t$ becomes fixed. In fact, it is possible to derive from the
invariants in Section~\ref{sc:formal}, that this order is the same one
$t$ was assigned at registration, \ie, the linearization point of
this write is line~\lineWrtWrt.

In line~\lineWrtFnz, $\aux{finalize}$ moves the write event $t$ from
the joint history $\histJ$ to the thread's self history $\histS$, thus
acknowledging that $t$ has terminated. The currently largest timestamp
of $\hist$ is recorded in $\E$ as $t$'s ending time. By definition of
$\stableorder$, all the writes that terminated before $t$ in real
time, will be ordered before $t$ in $\stableorder$.

\subparagraph*{Auxiliary code for \jyscan.} 
Method $\aux{set}$ toggles the scanner state $\sss$ on and off. When
executed in line~\lineScanUnsetsS, it returns the timestamp $\toff$
that is currently maximal in real time, as the moment when the scanner
is turned off.
%

The procedure $\aux{clear}(p)$ is executed in
lines~\lineScanClearsX--\lineScanClearsY~simultaneously with clearing
the forwarding pointer for $p$. In addition to recording that the
scanner passed lines~\lineScanClearsX\ or
respectively~\lineScanClearsY, by setting the $\spp$ bit, it colors
the subhistory $\histp$ green. Thus, by definition of
$\scanned{\stableorder}$, the ongoing one and all previous writes to
$p$ are recorded as scanned, and thus linearized.

Finally, the key auxiliary procedure of our approach is
$\aux{relink}$. It is executed at line~\lineScanRelinks~just before
the scanner returns the pair $(r_x, r_y)$. Its task is to modify the
logical order of the writes, to make $(r_x, r_y)$ \emph{appear} as a
valid snapshot. This will always be possible under the precondition of
$\aux{relink}$ that the timestamps $t_x$, $t_y$ of the events that
wrote $r_x$, $r_y$ respectively, are either the last green or the
yellow ones in the respective histories $\histx$ and $\histy$, and
$\aux{relink}$ will consider all four cases.  This precondition holds
after line~\lineScanChoosesRY~in Figure~\ref{fig:fcsl-snapshot}, as
one can prove from Invariants~\ref{inv:color} and~\ref{inv:readFP}. In
the precondition we introduce the following abbreviation:
\begin{equation}\label{eq:lgVy}
\hfill \lgVy\ p\, t\, \eqdef t = \mathsf{last\_green}_{\ordlist}\,
       \histp \vee \C(t) = \mathsf{yellow}\hfill
\end{equation}
%

$\aux{Relink}$ uses two helper procedures $\aux{inspect}$ and
$\aux{push}$, to change the logical order. $\aux{Inspect}$ decides if
the selected $t_x$ and $t_y$ determine a valid snapshot, and
$\aux{push}$ performs the actual reordering. The snapshot determined
by $t_x$ and $t_y$ is valid if there is no event $s$ such that
$t_x \tle s \tle t_y$ and $s$ is a write to $x$ (or, symmetrically
$t_y \tle s \tle t_x$, and $s$ is a write to $y$). If such $s$ exists,
$\aux{inspect}$ returns $\mathsf{Yes}\ x\ s$ (or $\mathsf{Yes}\ y\ s$
in the symmetric case). The reordering is completed by $\aux{push}$,
which moves $s$ right after $t_y$ (after $t_x$ in the symmetric case)
in $\tleq$. Finally, $\aux{relink}$ colors $t_x$ and $t_y$ green, to
fix them in $\stableorder$. We can then prove that $(r_x, r_y)$ is a
valid snapshot wrt.~$\stableorder$, and remains so under interference.
Notice that the timestamp $s$ returned by $\aux{inspect}$ is always
uniquely determined, and yellow. Indeed, since $t_x$ and $t_y$ are not
red, no timestamp between them can be red either
(Invariant~\ref{inv:redzone}). If $t_x \tle s \tle t_y$ and $s$ is a
write to $x$ (and the other case is symmetric), then $t_x$ must be the
last green in $\histx$, forcing $s$ to be the unique yellow timestamp
in $\histx$, by Invariant~\ref{inv:color}.

To illustrate, in Figure~\ref{fig:reorder:before} we have $r_x = 2$,
$r_y = 1$, $t_x$ and $t_y$ are both the last green timestamp of
$\histx$ and $\histy$, respectively, and $t_x \tle t_y$. However,
there is a yellow timestamp $s$ in $\histx$ coming after $t_x$,
encoding a write of $3$. Because $t_x \tle s \tle t_y$, the pair
$(r_x, r_y)$ is not a valid snapshot, thus $\aux{inspect}$ returns
$\mathsf{Yes}\ x\ s$, after which $\aux{push}$ moves $3$ after $1$.

%


We have omitted the definitions of $\aux{inspect}$ and $\aux{push}$
for the sake of brevity. These are presented in
Appendix~\ref{sc:relink-lemmas}. We conclude this section with the
main property of $\aux{relink}$, whose proof can be found in our Coq
files~\cite{CoqFiles}.

\begin{lemma}[Main property of $\aux{relink}$]\label{lem:relink-prefix}
Let the precondition of $\aux{relink}$ hold, \ie, $\sss = \sOff(\_)$,
$t_x \hpts (x, r_x), t_y \hpts (y, r_y) \in \hist$, $\sx = \sy =
\TT$, and $\forall p \in \{x,y\}\ldot \lgVy\ p\ t_p$. Then the ending
state of $\aux{relink}$ satisfies the following:
 \begin{enumerate}
 \item\label{lem:relink-lgVy} For all $p \in \{x, y\}$, $t_p =
   \mathsf{last\_green}_{\ordlistP}\, \histp'$.
 \item\label{lem:relink-green} Let $t = \mathsf{max}_{\ordlistP}
   (t_x,t_y)$. Then for every $s \leq_{\ordlistP}t$, $\CP(s) = \mathsf{green}$.
 \end{enumerate}
\end{lemma}

\section{Correctness}
\label{sc:proof}

\def\botLGY{{\ensuremath{\mathsf{fwdLastGY}}}}
\def\histLGY{{\ensuremath{\mathsf{lastGYHist}}}}
\def\greenH{{\ensuremath{\mathsf{green\_prefix}}}}
\newcommand{\spz}{S_z}

\begin{figure}[t]
\[
\begin{array}[t]{r l}
  \num{1} & \specK{\{ \histS = \hempty  \wedge
                      w \subseteq \stableorder \wedge h \subseteq \hist \wedge h_o \subseteq \histO \}} \\
  \num{2} & \specK{\{ \histS = \hempty \wedge \wpp =\wInit \wedge
                      w \subseteq \stableorder \wedge h \subseteq \hist \wedge h_o \subseteq \histO \}}\\
  \num{3} &\lat\,\actwrite{p}{v}; \aux{register}(v) \rat; \\
  \num{4} & \specK{\{\exists t\, \ldot\,
                      \histS = \hempty \wedge
                      \wpp = \wWrite\ t\ v \wedge t\hpts(p,v) \in \histJ \wedge
                      \dom{h_{\othersub}} \cup \scanned{w}
                      \subseteq \sideal{\stableorder}{t}\}}\\
  \num{5} & \lat\, b \tbnd \act{read}(S);\ \aux{check}(p,b) \rat;\\
  \num{6} & \specK{\{\exists t\ldot \histS = \hempty \wedge\!
               \wpp =  \textrm{if}\ b\
               \textrm{then}\ \wDirty\ t\ v\ \textrm{else}\ \wClean\ t\ v \wedge\ t\hpts(p,v) \in \histJ \wedge \hbox{}}\\
          & \specK{\hphantom{\{\exists t\ldot \hbox{}}
              \dom{h_{\othersub}} \cup \scanned{w}\!
                   \subseteq \sideal{\stableorder}{t} 
               \}}\\
  \num{7} & \kw{if}\ b\ \kw{then}\ \lat\, \actwrite{\aleksfwdp{p}}{v};\ \aux{forward}(p,v)\rat;\\
  \num{8} & \specK{\{\exists t\ldot \histS = \hempty \wedge
                \wpp = \wClean\ t\ v \wedge t\hpts(p, v) \in \histJ \wedge
                \dom{h_{\othersub}} \cup \scanned{w} \subseteq \sideal{\stableorder}{t} \}}\\
  \num{9} & \lat\,\aux{finalize}(i,v)\rat \\
  \num{10} & \specK{\{\exists t\ldot \histS = t \hpts (p,v) \wedge
              \dom{h_{\othersub}} \cup \scanned{w} \subseteq \sideal{\stableorder}{t} \}}
\end{array}
\]
  \caption{\label{proof:write} Proof outline for $\jywrite$.}
\end{figure}

We can now show that \jywrite\ and \jyscan\ satisfy the specifications
from Figure~\ref{fig:specs}.  As before, we avoid VDM notation in
proof outlines by using logical variables.
%
%
%
%
%
%

\subparagraph*{Proof outline for \jywrite}
The proof outline for \jywrite is presented in
Figure~\ref{proof:write}.
Line 1 introduces logical variables $w$, $h$ and $h_o$, which name the
initial values of $\stableorder$, $\hist$, and $\histO$.
Line 2 adds the knowledge that the writer for the pointer $p$ is
turned off ($\wpp = \wInit$). This follows from our implicit
assumption that there is only one writer in the system, which, in the
Coq code, we enforce by locks.

Line 3 is the first command of the program, and the most important
step of the proof. Here $\aux{register}$ allocates a fresh timestamp
$t$ for the write event, puts $t$ into $\histJ$, coloring it yellow or
red, and changes $\wpp$ to $\wWrite\ t\ v$, simultaneously with the
physical update of $p$ with $v$ (see Figure~\ref{fig:auxcode}). The
importance of the step shows in line 4, where we need to establish
that $t$ is placed into the logical order after all the other finished
or scanned events (\ie, $ \dom{h_\othersub} \cup \scanned\,
\stableorder \subseteq \sideal{\stableorder}{t}$). This information is
the most difficult part of the proof, but once established, it merely
propagates through the proof outline.

Why does this inclusion hold? From the definition, we know that
$\aux{register}$ appends $t$ to the end of the list $\ordlist$ (the
clause $\ordlistP = \mathsf{snoc}\ {\ordlist}\ t$ in the definition of
$\aux{register}$ in Figure~\ref{fig:auxcode}). Thus, after the
execution of line 3, we know that for every other timestamp $s$, $s
\tle t$. In particular, $s \neq t$, so it suffices to prove
$s\ {\stableorder}\ t$.
We consider two cases: $s \in \dom{h_o}$ and $s \in
\scanned{\stableorder}$.  In the first case, by
Invariant~\ref{inv:dom-tau}, $s \in \dom{\E}$. By freshness of $t$
wrt.~global history $h$ (which includes $h_o$), we get $\E(s) < t$,
and then the desired $s\ {\stableorder}\ t$ follows from the
definition of $\stableorder$.  In the second case, by definition of
$\scanned$, $\C(s) = \mathsf{green}$. Since $s \tle t$, the result
again follows by definition of $\stableorder$.

Still regarding line 4, we note that $t \in \dom{\histJ}$ holds
despite the interference of other threads. This is ensured by the
Invariant~\ref{inv:joint-hist}, because no other thread but the writer
for $p$, can modify $\wpp$. Thus, this property will continue to hold
in lines 6 and 8.

In line 6, the writer state $\wpp$ is updated following the definition
of the auxiliary procedure $\aux{check}$. The conjunct on $\dom{h_o}
\cup \scanned{w} \subseteq \sideal{\stableorder}{t}$ propagates from
line 4, by monotonicity of $\stableorder$ (Invariant~\ref{inv:mono}).
Similarly, in line 8, $\wpp$ is changed following the definition of
$\aux{forward}$, and the the other conjunct
propagates. $\aux{Forward}$ further colors a number of timestamps
green, but this is done in order to satisfy the state space invariants
from Section~\ref{sc:formal}, and is not exposed in the proof of
\jywrite.
Finally, in line 10, $\aux{finalize}$ moves $t\mapsto(p, v)$ from
$\histJ$ to $\histS$, thus completing the proof.




\begin{figure}[!htp]
\[
  \begin{array}[t]{r l}
  \num{1} & \specK{\{ \histS = \hempty \wedge h \subseteq \hist \}} \\
  \num{2} & \specK{\{ \histS = \hempty \wedge
                \sss = \sOff\, \_ \wedge \sx = \sy = \FF
                \wedge h \subseteq \hist\}}\\
  \num{3} & \lat\,\actwrite{\s}{\esc{true}};\ \aux{set}(\esc{true}) \rat;\\
  \num{4} & \specK{\{\, \histS = \hempty \wedge
              \sss = \sOn \wedge \sx = \sy = \FF \wedge
              h \subseteq \hist \}}\\
  \num{5} & \lat\,\actwrite{\var{\fx}}{\bot};\ \aux{clear}(x)\rat;\\
  \num{6} & \specK{\{
                 \histS = \hempty \wedge \sss= \sOn \wedge \sx = \TT
                 \wedge \sy = \FF \wedge
               h \subseteq \hist \wedge
               \C(\dom{h_{x}}) = \mathsf{green}\}}\\
  \num{7} & \lat\,\actwrite{\var{\fy}}{\bot};\ \aux{clear}(y)\rat;\\
  \num{8} & \specK{\{\histS = \hempty \wedge \sss= \sOn \wedge
             \sx = \sy = \TT \wedge
             h \subseteq \hist \wedge \C(\dom{h}) = \mathsf{green}\}}\\
  \num{9} & \var{vx} \tbnd \lat \act{read}(x) \rat;\\
  \num{10} & \specK{\{ \exists\, t_x\ldot\, \histS = \hempty \wedge
                  \sss = \sOn \wedge \sx = \sy = \TT  \wedge}\\
           & \specK{\hphantom{\{ \exists\, t_x\ldot}
                h \subseteq \hist \wedge \C(\dom{h}) = \mathsf{green} \wedge    
                \botLGY\ x\, t_x\, \var{vx} \}} \\
  \num{11} & \var{vy} \tbnd \lat \act{read}(y) \rat;\\
  \num{12} & \specK{\{ \exists\, t_x\, t_y\ldot\, \histS = \hempty \wedge
              \sss = \sOn \wedge \sx = \sy = \TT  \wedge
              h \subseteq \hist \wedge}\\
           & \specK{\hphantom{\{ \exists\, t_x \, t_y\ldot}
             \C(\dom{h}) = \mathsf{green} \wedge    
            \botLGY\ x\, t_x\, \var{vx} \wedge \botLGY\ x\, t_x\, \var{vy}\}} \\
  \num{13} &\lat\,\actwrite{\s}{\esc{false}};\ \aux{set}(\esc{false}) \rat;\\\
  \num{14} & \specK{\{ \exists\, t_x\, t_y\, \toff \ldot\, \histS = \hempty \wedge
             \sss = \sOff\, \toff \wedge \sx = \sy = \TT  \wedge
             h \subseteq \hist \wedge}\\
           & \specK{\hphantom{\{ \exists\, t_x \, t_y\, \toff\ldot}
             \C(\dom{h}) = \mathsf{green} \wedge    
            \botLGY\ x\, t_x\, \var{vx} \wedge \botLGY\ y\, t_y\, \var{vy}\}} \\
  \num{15} & \var{ox} \tbnd \lat \act{read}(\var{\fx}) \rat;\\
  \num{16} & \specK{\{\, \exists\, t_y\, t'_x\, \toff \ldot\,
              \histS = \hempty \wedge
              \sss = \sOff\ \toff \wedge \sx = \sy = \TT \wedge
              h \subseteq \hist \wedge}\\
           & \specK{\hphantom{\{\, \exists\, t_y\, t'_x\, \toff \ldot}\,
              \C(\dom{h}) = \mathsf{green} \wedge
              \botLGY\ y\, t_y\, \var{vy} \wedge}\\
           & \specK{\hphantom{\{\, \exists\, t_y\, t'_x\, \toff \ldot}\,
              \histLGY\ x\ t'_x\ (\textrm{if}\ r = \bot\
                   \textrm{then}\ \var{vx}\ \textrm{else}\, r)\}}\\
  \num{17} & \var{oy} \tbnd \lat \act{read}(\var{\fy}) \rat;\\
  \num{18} & \specK{\{\, \exists\, t'_x\, t'_y\, \toff \ldot\,
              \histS = \hempty \wedge
              \sss = \sOff\ \toff \wedge \sx = \sy = \TT \wedge
              h \subseteq \hist \wedge}\\
             & \specK{\hphantom{\{\,
                  \exists\, t'_x\, t'_y\, \toff \ldot}\,
              \C(\dom{h}) = \mathsf{green} \wedge
              \histLGY\ x\ t'_x\
                      (\textrm{if}\ \var{ox} = \bot\
                       \textrm{then}\ \var{vx}\
                       \textrm{else}\, \var{ox}) \wedge}\\
             & \specK{\hphantom{\{\,
                  \exists\, t'_x\, t'_y\, \toff \ldot}\,
                 \histLGY\ y\ t'_y\
                       (\textrm{if}\ \var{oy} = \bot\
                        \textrm{then}\ \var{vy}\ \textrm{else}\, \var{oy})\}}\\
  \num{19} & \var{rx} \tbnd \kw{if}\ (\var{ox} \neq\bot)\
                \kw{then}\ \var{ox}\ \kw{else}\ \var{vx};\\
  \num{20} & \var{ry} \tbnd \kw{if}\ (\var{oy} \neq\bot)\
                 \kw{then}\ \var{oy}\ \kw{else}\ \var{vy};\\
   \num{21} & \specK{\{\, \exists\, t'_x\, t'_y\, \toff \ldot\,
              \histS = \hempty \wedge
              \sss = \sOff\ \toff \wedge \sx = \sy = \TT \wedge
              h \subseteq \hist \wedge}\\
             & \specK{\hphantom{\{\,
                  \exists\, t'_x\, t'_y\, \toff \ldot}\,
              \C(\dom{h}) = \mathsf{green} \wedge
              \histLGY\ x\ t'_x\ \var{rx} \wedge
              \histLGY\ y\ t'_y\ \var{ry}\}}\\
  \num{22} & \lat\,\aux{relink}(\var{rx}, \var{ry});\
                \kw{return}\ (\var{rx}, \var{ry})\,\rat \\
  \num{23} & \specK{\{\, r\ldot \exists t\ldot \histS = \hempty \wedge
    r = \eval\ t\ {\stableorder}\ {\hist} \wedge
    \dom{h} \subseteq \ideal{\stableorder}{t} \wedge
    t \in \scanned{\stableorder}\}}
\end{array}
\]
  \caption{\label{proof:scan} Proof outline for $\jyscan$.}
\end{figure}

\subparagraph{Proof outline for \jyscan.}
Finally, the proof outline for is given in
Figure~\ref{proof:scan}. Line 1 introduces the logical variable $h$ to
name the initial $\hist$. Line 2 adds the knowledge that $\sss =
\sOff\, \_$ and $\sx = \sy = \FF$, \ie, that there are no other
scanners around, which is enforced by locking in our Coq files.

Line 3 is the first line of the code; it simply sets the scanner bit
$S$, and the auxiliaries $\sx$ and $\sy$, following the definition of
$\aux{set}$. The conjunct $h \subseteq \hist$ follows from
monotonicity by Invariant~\ref{inv:mono}.
The first important property comes from the lines 5 and 7. In these
lines, $\aux{clear}$ sets the values of $\sx$ and $\sy$, but,
importantly, also colors the events from $h$ green, first coloring
$x$-events, and then $y$-events. This will be important at the end of
the proof, where the fact that $h$ is all green will enable inferring
the postcondition. Moreover, because green events are never
re-colored, we propagate this property to subsequent lines without
commentary.

The read from $\x$ in line 9, and from $\y$ in line 11, must return
the last green, or the yellow event of their pointer, if no values are
forwarded in $\fx$ and $\fy$, respectively. This holds by
Lemma~\ref{lemma:first-read}, and is reflected by the conjuncts
$\botLGY\ x\ t_x\ \var{vx}$ and $\botLGY\ x\ t_x\ \var{vy}$ in line
12, where:
\[
 \hfill \botLGY\ p\ t\ v \eqdef \aleksfwdp{p} \hpts \bot
 \implies \lgVy\ p\ t \wedge t \hpts (p, v) \in \hist \hfill
\]
The implication guard $\aleksfwdp{p} \hpts \bot$ will be stripped away
in the future, if and when the reads of forwarding pointers in lines
15 and 17 observe that no forwarding values exist.

In line~13, the scanner unsets the bit $\s$ and records the ending
time of the scanner into the variable $\toff$ in line 14. The
conjuncts $\botLGY\ x\ t_x\ vx$ and $\botLGY\ y\ t_y\ vy$ from line 12
transfer to line 14 directly. This is so because $\aux{set}$ does not
change any colors. Moreover, any writes that may run concurrently with
this scan cannot invalidate the conjuncts. To see this, assume that we
had a concurrent \jywrite\ to $\x$ (reasoning is symmetric for $y$).
Such a \jywrite\ may add a new yellow timestamp $s$, but only if $t_x$
itself is the last green, in accord with Invariant~\ref{inv:color}. In
that case, $t_x$ remains the last green timestamp, and
$\botLGY\ x\ t_x\ vx$ remains valid. The concurrent \jywrite\ may
change the color of $s$ to green, by invoking $\aux{forward}$
(Figure~\ref{fig:fcsl-snapshot}, line 5), but then $\fx$ becomes
non-$\bot$, thus making $\botLGY\ x\ t_x\ vx$ hold trivially.

In lines~15 and~17, \jyscan\ reads from the forwarding pointers $\fx$
and $\fy$ and stores the obtained values into $ox$ and $oy$,
respectively.  By Invariant~\ref{inv:readFP}, we know that if $ox \neq
\bot$, there exists $t'_x$ s.t. $t'_x \hpts (x,ox) \in \hist$, and
$t'_x$ is the last green or yellow write event of $\hist_x$.  In case
$ox = \bot$, we know from the $\botLGY$ conjunct preceding the read
from $\fx$, that such last green or yellow event is exactly $t_x$.
The consideration for $\fy$ is symmetric, giving us the assertion in
line 18, where:
\[
\hfill\histLGY\ p\ t\ v \eqdef
\lgVy\ p\ t \wedge t \hpts (p, v) \in \hist\hfill
\]

Next, line~19 merely names by $rx$ the value of $vx$, if $ox$ equals
$\bot$, and similarly for $ry$ in line 20, leading to
line~21. Finally, on line~22, the method finishes by invoking
$\lat\,\aux{relink}(\var{rx}, \var{ry});\ \kw{return}\ (\var{rx},
\var{ry})\,\rat$. Thus, it returns the selected snapshot $(r_x,r_y)$
and relinks the events so that the $\stableorder$ justifies the choice
of snapshots.



We prove that the final state satisfies the postcondition in line 23,
by using the main property of $\aux{relink}$
(Lemma~\ref{lem:relink-prefix}).
First, we pick $t = \mathsf{max}_{\ordlist}(t'_x, t'_y)$. Then $r =
\eval\ t\ \stableorder\ \hist$ holds, by the following argument. By
Lemma~\ref{lem:relink-prefix}.\ref{lem:relink-lgVy}, $rx$ is the value
of the last green timestamp in $\histx$. By
Lemma~\ref{lem:relink-prefix}.\ref{lem:relink-green}, all the
timestamps below $t$ are green, thus $rx$ is the value of the
\emph{last} timestamp in $\histx$ that is smaller or equal to $t$.  By
a symmetric argument, the same holds of $ry$. But then, the pair $r =
(rx, ry)$ is the snapshot at $t$, \ie, equals
$\eval\ t\ \stableorder\ \hist$.

The conjunct $t \in \scanned{\stableorder}$ is proved as
follows. Unfolding the definition of $\scanned$, we need to show
$\ideal{\stableorder}{t} = \ideal{{\tleq}}{t}$, and $\forall s\in
\ideal{\stableorder}{t}\ldot \C(s) = \mathsf{green}$.  The first
conjunct follows from Lemma~\ref{lemma:complete-green}. The second
immediately follows from the first by
Lemma~\ref{lem:relink-prefix}.\ref{lem:relink-green}.

To establish $\dom{h} \subseteq \ideal{\stableorder}{t}$, we proceed
as follows. Let $s \in \dom{h}$. From line 21, we know $\C(s) =
\mathsf{green}$. Because $t'_x$ and $t'_y$ are last green (by
$\ordlist$) or yellow events, by Invariant~\ref{inv:color} it must be
$s \tleq t'_x, t'_y$, and thus $s \tleq t$. However, we already showed
that $\ideal{\stableorder}{t} = \ideal{{\tleq}}{t}$. Thus,
$s\ {\stableorder}\ t$, finally establishing the postcondition.

\section{Discussion}
\label{sc:discussion}

\subparagraph*{Comparison with linearizability, revisited}
  
As we argued in Section~\ref{sc:formal}, our specifications for the
snapshot methods directly capture that the method calls can be placed
in a linear sequence, in a way that preserves the order of
non-overlapping calls. This is precisely what linearizability achieves
as well, but by technically different means. We here discuss some
similarities and differences between our method and linearizability.

The first distinction is that linearizability is a property of a
concurrent object, whereas our specifications are ascribed to
individual methods, as customary in Hoare logic. This immediately
enables us to use an of-the-shelf Hoare logic, such as FCSL, for
specification. 

Second, linearizability draws its power from the connection to
contextual refinement~\cite{FilipovicOHRW+TCS10}: one can substitute a
potentially complex method $A$ in a larger context, by a simpler
method $B$, to which $A$ linearizes. In our setting, such a property
is enabled by a general substitution principle, which says that
programs with the same spec can be interchanged in a larger context,
without affecting the larger context's proof. Moreover, contextual
refinement (and thus linearizability) is defined for general programs,
without regard to their preconditions and postconditions. However, it
is often the case that the refinement only holds if the substituted
programs satisfy some Hoare logic spec. In this sense, our setting is
more expressive, since the substitution principle is given relative to
a Hoare logic spec.

Finally, while our specification of the snapshot methods are motivated
by linearizability, there is no requirement---and hence no
proof---that an FCSL specification implies linearizability. But this
is a feature, rather than a bug. It enables us to specify and combine,
in one and the same logic, programs that are linearizable, with those
that are not. We refer to~\cite{SergeyNBD+OOPSLA16} for examples of
how to specify and verify non-linearizable programs in FCSL.

\begin{wrapfigure}[12]{t}{0.4\textwidth} 
%
%
\begin{minipage}[t]{.4\textwidth}
\[\hfill
\begin{array}{rl}
\num{1} & \esc{scan}\ () : ( A \times A )~ \{ \\ 
\num{2} & ~~~ (\var{cx}, \var{vx}) \tbnd \act{read}(\x);\\
\num{3} & ~~~ (\var{cy}, \_ ) \tbnd \act{read}(\y);\\
\num{5} & ~~~ (\_ , \var{tx}) \tbnd \act{read}(\x);\\
\num{5} & ~~~ \kw{if}\ vx = tx \\
\num{6} & ~~~ \kw{then}\ \kw{return}\ (\var{cx},\var{cy})\\
\num{7} & ~~~ \kw{else}\ \esc{scan}\ (); \}
\end{array}\hfill
\]
\end{minipage}
%
%
\caption{{\tt scan} using versions.}
\label{fig:readpair}
\end{wrapfigure}



\subparagraph*{Alternative snapshot implementations.}
FCSL's substitution principle can be exploited further in an
orthogonal way: it allows us to re-use the specs for \jywrite\ and
\jyscan\ in Figure~\ref{fig:specs}, ascribing them to a different
concurrent snapshot algorithm. For that matter, we re-visit the
previous verification in FCSL of the pair-snapshot
algorithm~\cite{SergeyNB+ESOP15}. We present only \jyscan\ in
Figure~\ref{fig:readpair}, as \jywrite\ is trivial.

In this example, the snapshot structure consists of pointers $x$ and
$y$ storing tuples $(c_x, v_x)$ and $(c_y, v_y)$, respectively. $c_x$
and $c_y$ are the payload of $x$ and $y$, whereas $v_x$ and $v_y$ are
version numbers, internal to the structure. Writes to $x$ and $y$
increment the version number, while {\tt scan} reads $x$, $y$ and $x$
again, in succession. Snapshot inconsistency is avoided by restarting
if the two version numbers of $x$ differ. In this paper's notation,
the specification proved for \jyscan in~\cite{SergeyNB+ESOP15} reads:
\[\hfill
\esc{scan} : \tsPos{\{\histS = \hempty\}}\
\tsPos{\{\exists t\ldot \histS' = \hempty \wedge
  r = \eval\ t\ \histP \wedge \dom{\hist}
  \subseteq \ideal{\histP}{t}\}}\hfill
\]
This spec is indeed very similar to the one of \jyscan in
Figure~\ref{fig:specs}, but exhibits that the algorithm does not
require dynamic modification to the event ordering. Thus, by defining
$\stableorder$ to be the natural ordering on timestamps in the global
history $\hist$ (so that $\ideal{\stableorderP}{t} =
\ideal{\histP}{t}$), and taking $\scanned{\stableorder}$ to be the set
of all timestamps in $\hist$ (so that $t \in \scanned{\stableorder}$
is trivially true and can be added to the postcondition above), the
above spec directly weakens into that of Figure~\ref{fig:specs}.
Since client proofs are developed in FCSL out of the specs, and not
the code of programs, we can substitute different implementations of
snapshot algorithms in clients, without disturbing the clients'
proofs.
This is akin to the property that programs that linearize to the same
sequential code are interchangeable in clients.

\subparagraph*{Relation to Jayanti's original proof.}
\label{sec:relat-jayant-orig}
Finally, we close this section by noting that our proof of Jayanti's
algorithm seems very different from Jayanti's original proof. Jayanti
relies on so-called \emph{forwarding principles}, as a key property of
the proof. For example, Jayanti's First Forwarding Principle says (in
paraphrase) that if {\tt scan} misses the value of a concurrent write
through lines~\lineScanReadsX--\lineScanReadsY\ of
Figure~\ref{fig:jayanti-snapshot}, but the write terminates before the
scanner goes through line~\lineScanUnsetsS\ (the linearization point
of \jyscan), then the scanner will catch the value in the forwarding
pointers through lines~\lineScanReadsFX--\lineScanReadsFY.
Instead of forwarding principles, we rely on colors to algorithmically
construct the status of each write event as it progresses through
time, and express our assertions using formal logic. For example,
though we did not use the First Forwarding Principle, we nevertheless
can express a similar property, whose proof follows from
Invariants introduced in Section~\ref{sc:auxiliaries}:
\begin{proposition}\label{inv:fwd1}%
If $\sss = \sOff\ \toff$ and $\sx = \sy = \TT$---i.e., the scanner is
in lines~\lineScanReadsFX--\lineScanChoosesRY\ and it has unset $\s$
in line~\lineScanUnsetsS\ at time $\toff$---then: $ \forall t \in
\hist\ldot\ t \leq \E(t)< \toff \implies \C(t)= \mathsf{green}$.
\end{proposition}


\section{Related work}
\label{sc:related}


\subparagraph*{Program logics for linearizability.}

The proof method for establishing linearizability of concurrent
objects based on the notion of linearization points has been presented
in the original paper by Herlihy and
Wing~\cite{HerlihyW+TOPLAS90}. The first Hoare-style logic, employing
this method for compositional proofs of linearizability was introduced
in Vafeiadis' PhD thesis~\cite{VafeiadisHHS+PPoPP06, Vafeiadis+PhD07}.
However, that logic, while being inspired by the combination of
Rely-Guarantee reasoning and Concurrent Separation
logic~\cite{VafeiadisP+CONCUR07} with syntactic treatment of
linearization points~\cite{VafeiadisHHS+PPoPP06}, did not connect
reasoning about linearizability to the verification of client programs
that make use of linearizable objects in a concurrent environment.


%


Both these shortcomings were addressed in more recent works on program
logics for linearizability~\cite{LiangF+PLDI13,KhyzhaGP+FM16}, or,
equivalently, \emph{observational
  refinement}~\cite{FilipovicOHRW+TCS10,TuronDB+ICFP13}. These works
provided semantically sound methodologies for verifying refinement of
concurrent objects, by encoding atomic commands as resources
(sometimes encoded via a more general notion of
\emph{tokens}~\cite{KhyzhaGP+FM16}) directly into a Hoare
logic. Moreover, the logics~\cite{LiangF+PLDI13, TuronDB+ICFP13}
allowed one to give the objects standard Hoare-style specifications.
However, in the works~\cite{LiangF+PLDI13,TuronDB+ICFP13}, these two
properties (\ie,~linearizability of a data structure and validity of
its Hoare-style spec) are established separately, thus doubling the
proving effort.
That is, in those logics, provided a proof of linearizability for a
concurrent data structure, manifested by a spec that suitably handles
a \emph{command-as-resource}, one should then devise a declarative
specification that exhibits temporal and spatial aspects of executions
(akin to our history-based specs from Figure~\ref{fig:specs}),
required for verifying the client code.

Importantly, in those logics, determining the linearization order of a
procedure is tied with that procedure ``running'' the
command-as-resource within its execution span. This makes it difficult
to verify programs where the procedure terminates before the order is
decided on, such as \jywrite~operation in Jayanti's snapshot. The
problem may be overcome by extending the scope of \emph{prophecy
  variables}~\cite{AbadiL+lics88} or \emph{speculations} beyond the
body of the specified procedure. However, to the best of our
knowledge, this has not been done yet.


%

\subparagraph*{Hoare-style specifications as an alternative to
  linearizability.} 

A series of recent Hoare logics focus on specifying concurrent
behavior \emph{without} resorting to
linearizability~\cite{SergeyNB+ESOP15, SergeyNBD+OOPSLA16, SvendsenB+ESOP14,
  PintoDYG+ECOOP14, JungSSSTBD+POPL15}.
This paper continues the same line of thinking, building
on~\cite{SergeyNB+ESOP15}, which explored patterns of assigning
Hoare-style specifications with self/other auxiliary histories to
concurrent objects, including \emph{higher-order} ones (e.g., {flat
  combiner}~\cite{HendlerHIST+SPAA10}), and \emph{non-linearizable}
ones~\cite{SergeyNBD+OOPSLA16} in FCSL~\cite{NanevskiLSD+ESOP14}, but
has not considered non-local, future-dependent linearization points,
as required by Jayanti's algorithm.
%

Alternative logics, such as
Iris~\cite{JungSSSTBD+POPL15,JungKBD+ICFP16} and
iCAP~\cite{SvendsenB+ESOP14}, employ the idea of ``ghost
callbacks''~\cite{Jacobs-Piessens+POPL11}, to identify precisely the
point in code when the callback should be invoked.  Such a program
point essentially corresponds to a local linearization
point. Similarly to the logical linearizability proofs, in the
presence of future-dependent LPs, this method would require
speculating about possible future execution of the callback, just as
commented above, but that requires changes to these logics'
metatheory, in order to support speculations, that have not been
carried out yet.

The specification style of TaDA logic~\cite{PintoDYG+ECOOP14} is
closer to ours in the sense that it employs \emph{atomic tracking
  resources}, that are reminiscent of our history entries. However,
the metatheory of TaDA does not support ownership transfer of the
atomic tracking resources, which is crucial for verifying algorithms
with non-local linearization points. As demonstrated by this paper and
also previous works~\cite{SergeyNB+ESOP15,SergeyNBD+OOPSLA16}, history
entries can be subject to ownership transfer, just like any other
resources.


The key novelty of the current work with respect to previous results
on Hoare logics with histories~\cite{FuLFSZ+CONCUR10, LiangF+PLDI13,
  GotsmanRY+ESOP13, BellAW+SAS10, SergeyNB+ESOP15, HemedRV+DISC15} is
the idea of representing logical histories as auxiliary state, thus
enabling constructive reasoning, by \emph{relinking}, about
dynamically changing linearization points.
Since relinking is just a manipulation of otherwise standard auxiliary
state, we were able to use FCSL \emph{off the shelf}, with no
extensions to its metatheory. Furthermore, we expect to be able to use
FCSL's higher-order features to reason about higher-order (\ie,
parameterized by another data structure) snapshot-based
constructions~\cite{PetrankT+DISC13}.
Related to our result, O'Hearn \etal have shown how to employ
history-based reasoning and Hoare-style logic to
\emph{non-constructively} prove the existence of linearization points
for concurrent objects out of the data structure
invariants~\cite{OHearnRVYY+PODC10}; this result is known as \emph{the
  Hindsight Lemma}. The reasoning principle presented in this paper
generalizes that idea, since the Hindsight Lemma is only applicable to
``pure'' concurrent methods (\eg, a concurrent set's
\texttt{contains}~\cite{HellerHLMSS+OPODIS05}) that do not influence
the position of other threads' linearization points. In contrast, our
history relinking handles such cases, as showcased by Jayanti's
construction, where the linearization point of \texttt{write} depends
on the (future) outcome of \texttt{scan}.


\subparagraph*{Semantic proofs of linearizability.}
\label{sec:semant-proofs-line}

There has been a long line of research on establishing linearizability
using forward-backwards
simulations~\cite{SchellhornWD+CAV12,ColvinGLM06,ColvinDG05}. These
proofs usually require a complex simulation argument and are not
modular, because they require reasoning about the entire data
structure implementation, with all its methods, as a monolithic STS.

Recent works~\cite{HenzingerSV+CONCUR13, ChakrabortyHSV+LMCS15,
  DoddsHK+POPL15} describe methods for establishing linearizability of
sophisticated implementations (such as the Herlihy--Wing
queue~\cite{HerlihyW+TOPLAS90} or the time-stamped
stack~\cite{DoddsHK+POPL15}) in a modular way, via
\emph{aspect-oriented} proofs.
This methodology requires devising, for each class of objects (\eg,
queues or stacks), a set of specification-specific conditions, called
{\it aspects}, characterizing the observed executions, and then
showing that establishing such properties implies its linearizability.
This approach circumvents the challenge of reasoning about
future-dependent linearization points, at the expense of (a)
developing suitable aspects for each new data structure class and
proving the corresponding ``aspect theorem'', and (b) verifying the
aspects for a specific implementation. 
%
%
Even though some of the aspects have been mechanized and proved
adequate~\cite{DoddsHK+POPL15}, 
%
currently, we are not aware of such aspects for snapshots.



Our approach is based on program logics and the use of STSs to
describe the state-space of concurrent objects. Modular reasoning is
achieved by means of separately proving properties of specific STS
transitions, and then establishing specifications of programs,
composed out of well-defined atomic commands, following the
transitions, and respecting the STS invariants.

\subparagraph*{Proving linearizability using partial orders.}
Concurrently with us, Khyzha \etal~\cite{KhyzhaGP+ESOP17} have
developed a proof method for proving linearizability, which can handle
certain class of data structures with similar future dependent
behavior.
The method works by introducing a partial order of events for the data
structure as auxiliary state, which in turn defines the abstract
histories used for satisfying the sequential specification of the data
structure. Relations are added to this partial order at {\it
  commitment points} of the instrumented methods, which the verifier
has to identify.

The ultimate goal of this method is to assert the
linearizability of a concurrent data structure. As we have shown in
Section~\ref{sc:clients}, FCSL goes beyond as it provides a logical
framework to carry out formal proofs about the correctness of a
concurrent data structure and its clients.


The proof technique also tracks the ordering of events 
differently from ours. Where we keep a single witness for
the current total ordering of events at all stages of execution,
their technique requires keeping many witnesses. Their main theorem requires a
proof that all linearizations of the abstract histories---\ie all
possible linear extensions of the partial order into a total
order---satisfy the sequential specification of the data structure.


Through personal communication we learned that the
technique cannot apply, for instance, to the verification of the {\it
  time-stamped} (TS) stack~\cite{DoddsHK+POPL15}. This is because a
partial order does not suffice to characterize the abstract histories
required to verify the data structure.
In contrast, given the flexibility of FCSL in designing and reasoning
with auxiliary state, we believe that our technique would not suffer
such shortcomings.

\section{Conclusions}
\label{sec:conclusions}

The paper illustrates a new approach allowing one to specify that the
execution history of a concurrent data structure can be seen as a
\emph{sequence of atomic events}. The approach is thus similar in its
goals to linearizability, but is carried out exclusively using a
separation-style logic to uniformly represent the state and time
aspects of the data structure and its methods.

Reasoning about time using separation logic is very effective, as it
naturally supports \emph{dynamic and in-place updates} to the temporal
ordering of events, much as separation logic supports dynamic and
in-place updates of spatially linked lists. The need to modify the
ordering of events frequently appears in linearizability proofs, and
has been known to be tricky, especially when the order of a terminated
event depends on the future. In our approach, the modification becomes
a conceptually simple manipulation of auxiliary state of histories of
colored timestamps.




We have carried out and mechanized our proof of Jayanti's
algorithm~\cite{Jayanti+STOC05} in FCSL, without needing any additions
to the logic.
%
%
Such development, together with the fact that FCSL has previously been
used to verify a number of non-trivial concurrent
structures~\cite{SergeyNB+ESOP15,SergeyNB+PLDI15,SergeyNBD+OOPSLA16},
gives us confidence that the approach will be applicable, with minor
modifications, to other structures whose linearizations exhibit
dynamic dependence on the
future~\cite{DoddsHK+POPL15,Morrison-al:PPoPP13,Hoffman-al:OPODIS07}.


One modification that we envision will be in the design of the data
type of timestamped histories. In the current paper, a history of the
snapshot object needs to keep only the \jywrite\ events, but not the
\jyscan\ events. In contrast, in the case of stacks, a history would
need to keep both events for push and pop operations. But in FCSL,
histories are a \emph{user-defined} concept, which is not hardwired
into the semantics of the logic. Thus, the user can choose any
particular notion of history, as long as it satisfies the properties
of a Partial Commutative
Monoid~\cite{LeyWildN+POPL13,NanevskiLSD+ESOP14}. Such a history can
track pushes and pops, or any other auxiliary notion that may be
required, such as, \eg,~specific ordering constraints on the events.

\bibliography{references,procs}
\appendix
\section{A brief introduction to FCSL}
\label{sc:background}

A state of a resource in FCSL~\cite{NanevskiLSD+ESOP14}, such as that
of snapshot data structure discussed in this paper, always consists of
three distinct auxiliary variables that we name $a_\lcl$, $a_\env$ and
$a_\joint$. These stand for the abstract self state, other state, and
shared (joint) state.

However, the user can pick the types of these variables based on the
application. In this paper, we have chosen $a_\lcl$ and $a_\env$ to be
histories, and have correspondingly named them $\histS$ and $\histO$.
On the other hand, $a_\joint$ consists of all the other auxiliary
components that we discussed, such as the variables $\histJ$, $\E$,
$\C$, $\sx$, $\sy$, $\wx$ and $\wy$. These variables become merely
projections out of $a_\joint$.

It is essential that $a_\lcl$ and $a_\env$ have a common type, which
moreover, exhibits the algebraic structure of a \emph{partial
  commutative monoid} (PCM). A PCM requires a partial binary operation
$\bullet$ which is commutative and associative, and has a unit. 
%
%
PCMs are important, as they give a generic way to define the inference
rule for parallel composition.
\[
\hfill\begin{array}{c}
  e_1 : \specK{\{P_1\}}\ A\ \specK{\{Q_1\}} @ C \quad
  e_2 : \specK{\{P_2\}}\ B\ \specK{\{Q_2\}} @ C \\[2pt]
\hline\\[-7pt]
e_1 \parallel e_2 : \specK{\{P_1 \circledast P_2\}}\ (A \times B)\
\specK{\{[r.1/r]Q_1 \circledast [r.2/r]Q_2\}} @ C
\end{array}\hfill
\]
Here, $\circledast$ is defined over state predicates $P_1$ and $P_2$
as follows.
\[
\hfill\begin{array}{c}
  (P_1 \circledast P_2) (a_\lcl, a_\joint, a_\env) \iff
  \exists x_1~x_2\ldot\
  a_\lcl = x_1 \bullet x_2 \wedge
  P_1 (x_1, a_\joint, x_2 \bullet a_\env) \wedge
  P_2 (x_2, a_\joint, x_1 \bullet a_\env)
\end{array}\hfill
\]

The inference rule, and the definition of $\circledast$, formalize the
intuition that when a parent thread forks $e_1$ and $e_2$, then $e_1$
is part of the environment for $e_2$ and vice-versa. This is so
because the \emph{self} component $a_\lcl$ of the parent thread is
split into $x_1$ and $x_2$; $x_1$ and $x_2$ become the \emph{self}
parts of $e_1$, and $e_2$ respectively, but $x_2$ is also added to the
\emph{other} component $a_\env$ of $e_1$, and dually, $x_1$ is added
to the \emph{other} component of $e_2$.

In this paper, the PCM we chose is that of histories, which are a PCM
under the operation of disjoint union $\hunion$, with the $\hempty$
history as the unit. More common in separation logic is to use heaps,
which, similarly to histories, form PCM under disjoint (heap) union
and theempty heap, $\mathsf{empty}$. In FCSL, these can be combined
into a Cartesian product PCM, to enable reasoning about both space and
time in the same system.


The frame rule is a special case of the parallel composition rule,
obtained when $e_2$ is taken to be the idle thread.
\[
\hfill \begin{array}{c}
e : \specK{\{P\}}\ A\ \specK{\{Q\}} @ C\\[2pt]
\hline\\[-7pt]
e : \specK{\{P \circledast R\}}\ A\ \specK{\{Q \circledast R\}} @ C 
\end{array}  \quad \mbox{$R$ is stable} \hfill
\]

For the purpose of this paper, the rule is important because it allows
us to generalize the specifications of {\tt write} and {\tt scan} from
Figure~\ref{fig:specs}. In that figure, both procedures start with the
precondition that $\histS = \hempty$. But what do we do if the
procedures are invoked by another one which has already completed a
number of writes, and thus its $\histS$ is non-empty. By
$\circledast$-ing with the frame predicate $R \eqdef (\histS = k)$,
the frame rule allows us to generalize these specs into ones where the
input history equals an arbitrary $k$:
\[
\begin{array}{l}
\mathtt{write}\ (p, v) : 
\begin{array}[t]{l}
\tsPre{\{\histS = k\}}\\
\tsPos{\{\exists t\ldot \histS' = h \hunion t \mapsto (p, v) \wedge
    \dom {\histO} \cup \scanned\, \stableorder
       \subseteq \sideal{\stableorderP}{t}\}} @ C 
\end{array}\\[5pt]
\mathtt{scan} : 
\!\!\begin{array}[t]{l}
\tsPre{\{\histS = k\}}\\
\tsPos{\{r\ldot \exists t\ldot \histSP = k \wedge
   r =\! \eval\ t\, {\stableorderP}\, {\histP} \wedge
  \dom{\hist} \subseteq \ideal{\stableorderP}{t} \wedge
  t \in \scanned{\stableorderP}\}} @ C 
\end{array}
\end{array}
\]
These two {\it large-footprint} instances of the rules for {\jyscan}
and {\jywrite} are those used in the proof of our clients in
Section~\ref{sc:clients}. For further details on FCSL, its semantics
and implementation, we refer the reader to~\cite{NanevskiLSD+ESOP14}.

\section{Implementation and Correctness of $\aux{relink}$}
\label{sc:relink-lemmas}
\def\cat{{\mathbin{+\!\!+}}}

In Section~\ref{sc:implementation}, we described briefly the
implementation of $\aux{relink}$, without giving much details on the
auxiliary helper functions $\aux{inspect}$ and $\aux{push}$. We give
here their definitions, together with some associated properties:

\begin{definition}[inspect]\label{def:inspect}%
Given two timestamps $t_x$, $t_y$ then $\aux{inspect}\ t_x\, t_y\,
\ordlist\, \C$ is defined as follows:
\begin{equation*}
\aux{inspect}\ t_x\, t_y\ \C\ \eqdef
\begin{cases}
  \mathsf{Yes}\ x\ t_z\ \qquad \qquad
      \begin{array}[t]{l}
      \textrm{if}\ t_x \tle t_y,\ t_x = \mathsf{last\_green}_\ordlist\ \histx, \\
       t_z = \mathsf{yellow\_timestamp}_\ordlist\ \histx,\,
       \textrm{and} \ t_z \tle t_y
      \end{array} \\[5pt]
  \mathsf{Yes}\ y\ t_z\ \qquad \qquad
      \begin{array}[t]{l}
      \textrm{if}\ t_y \tle t_x,\ t_y = \mathsf{last\_green}_\ordlist\ \histy, \\
       t_z = \mathsf{yellow\_timestamp}_\ordlist\ \histy,\,
       \textrm{and} \ t_z \tle t_x
      \end{array} \\[5pt]
  \mathsf{No}\hphantom{\mathsf{s}\ x\ t_z}\ \qquad \qquad \textrm{otherwise}
\end{cases}
\end{equation*}
\end{definition}

\begin{definition}[push]\label{def:push}%
 $\aux{push}$ is a surgery operation defined on $\ordlist$ as
  follows:
\[\hfill
  \textrm{Let}\ \ordlist = \ordlist_{<_i}\ \cat
  \ i\ \cat\ \ordlist_{i..j}\ \cat\ j\ \cat\ \ordlist_{>_j},\
  \textrm{then}\
  \aux{push}\ i\ j\ \ordlist =
  \ordlist_{<_i} \cat\  \ordlist_{i..j}\ \cat\ j\ \cat\ i\ \cat \ordlist_{>_j}
  \hfill
\]
\end{definition}

The definition of $\aux{inspect}$ works under the assumption that
$t_x$ and $t_y$ are, respectively, the last green or yellow timestamp
in $\histx$ and $\histy$. This latter fact is recovered in the
definition of $\aux{relink}$ in Figure~\ref{fig:auxcode} and
reinforced in line~21 in the proof of {\jyscan} in
Figure~\ref{proof:scan}. When $\aux{inspect}$ returns
$\mathsf{Yes}\ p\, t_z$, $\ordlistP$ is computed by pushing some $i$
timestamp past another timestamp $j$ in $\ordlist$. The definition of
$\aux{push}$ above shows that this operation is an algebraic
manipulation on sequences. In fact, we implement it using standard
{\it surgery} operations on lists: $\cat$, $\mathsf{take}$, \etc.

In Section~\ref{sc:implementation}, we have mentioned that the
correctness aspect of auxiliary code involves proving that the code
preserves the auxiliary state invariants from
Section~\ref{sc:auxiliaries}. For example, the correctness proof of
$\aux{relink}$, relies on the following helper lemmas. The first lemma
asserts that $\aux{inspect}$ correctly determines the ``offending''
timestamp; the second and the third lemma assert that $\aux{push}$
modifies $\ordlist$ in a way that allows us to prove (in
Section~\ref{sc:proof}), that the pair $(r_x, r_y)$ a valid snapshot.

\begin{lemma}[Correctness of $\aux{inspect}$]\label{lem:inspect}
If $t_x$, $t_y$ are timestamps for write events of $r_x$, $r_y$, then
$\aux{inspect}\ t_x\ t_y\ \ordlist\ \C$ correctly determines that
$(r_x, r_y)$ is a valid snapshot under ordering $\tle$ and colors
$\C$, or otherwise returns the ``offending'' timestamp. More formally,
if $\sss = \sOff\ \toff, \sx =\TT, \sy =\TT$, and for each $p \in
\{x,y\}$, $ t_p \hpts (p, r_p) \in \hist$ and $\lgVy\ p\, t_p $, the
following are exhaustive possibilities.

\begin{enumerate}
 \item If $t_x \tle t_y$ and $ \C(t_x) = {\sf yellow}$, then
   $\aux{inspect}\ t_x\ t_y\ \ordlist\ \C =
   \mathsf{No}$. Symmetrically for $t_y \tle t_x$.

 \item If $ t_x \tle t_y $, $ t_x = \aux{last\_green}\ \histx$, and
       $\forall s \in \histx\ldot\ t_x \tle s\,{\implies}\,t_y \tle
       s$, then \\ $\aux{inspect}\ t_x\, t_y\, \ordlist\, \C
       = \mathsf{No}$. Symmetrically for $t_y \tle t_x$.

 \item If $ t_x <_{\ordlist} t_y $, $ t_x = \aux{last\_green}\ \histx
   $, $s \in \histx$, and $t_x \tle s \tle t_y$, it follows that
   $\aux{inspect}\ t_x\, t_y \ordlist \C = \mathsf{Yes}\, x\, s$ and
   $\C(s) = {\sf yellow}$. Symmetrically for $t_y \tle t_x$.
\end{enumerate}
\end{lemma}

\begin{lemma}[Push Mono]\label{lem:push-mono}
Given elements $a, b, i, j$, all in $\ordlist$, and $\ordlistP =
\aux{push}\ i\, j\, \ordlist$, then:
\begin{enumerate}
\item\label{lem:push:left} If $a \tle i$ then $ a \tle b \implies a
  \tleP b$.
\item\label{lem:push:right} If $j \tle b$ then $ a \tle b \implies a
  \tleP b $.
\item\label{lem:push:window} If $a \neq i$ then $ a \tle b \implies a
  \tleP b $
\end{enumerate}
\end{lemma}


\begin{lemma}[Correctness of $\aux{push}$]\label{lem:push}
Given $\sss = \sOff\ \toff, \sx = \sy =\TT$, and for $p \in \{x,y\}$,
we have $t_p \hpts (p, r_p) \in \hist$, $ \lgVy\ p\, t_p$, and
{$\aux{inspect}\ t_x\ t_y\ \ordlist\ \C = {\sf Yes}\ p\ t_s$}. If we
name $t_z \in \{ t_x, t_y\}$, with $p \neq z$, and $\ordlist' =
\aux{push}\ t_s\ t_z\ \ordlist$, then:
\begin{enumerate}
 \item $\aux{relink}$ satisfies the 2-state invariants from
   Invariant~\ref{inv:mono}.
\item $\histP, \ordlistP, \EP, \CP$ satisfies all the resource
  invariants from Section~\ref{sc:auxiliaries}, \ie
  Invariants~\ref{inv:overlap}--\ref{inv:dom-tau}.
\end{enumerate}
\end{lemma}

In our mechanization, these three lemmas allow us to prove
Lemma~\ref{lem:relink-prefix}, $\aux{relink}$'s main property.

\end{document}